\documentclass[a4paper,fleqn,usenatbib]{mnras}

\usepackage{mathptmx}

\usepackage[T1]{fontenc}
\usepackage{ae,aecompl}
\usepackage{url}

\usepackage{graphicx}	
\usepackage{amsmath}	
\usepackage{amssymb}	

\usepackage{booktabs}
\usepackage{threeparttable}
\usepackage{multirow}
\usepackage{amsmath}
\usepackage{bm}

\usepackage{diagbox}
\usepackage{booktabs}
\usepackage{threeparttable}
\usepackage[utf8]{inputenc}

\title[Critical spin periods of small cohesive asteroids]{Critical spin periods of sub-km-sized cohesive rubble-pile asteroids: dependencies on material parameters}

\author[S.C. Hu et al.]{
Shoucun Hu,$^{1, 2}$\thanks{E-mail: hushoucun@pmo.ac.cn}
Derek C. Richardson,$^{3}$
Yun Zhang,$^{4}$
Jianghui Ji$^{1, 2}$
\newauthor{
}
\\
$^{1}$CAS Key Laboratory of Planetary Sciences, Purple Mountain Observatory, Chinese Academy of Sciences, Nanjing 210033, China\\
$^{2}$CAS Center for Excellence in Comparative Planetology, Hefei, 230026, China\\
$^{3}$Department of Astronomy, University of Maryland, College Park, MD 20740-2421, USA\\
$^{4}$Université Côte d’Azur, Observatoire de la Côte d’Azur, CNRS, Laboratoire Lagrange, Nice, France
}
\date{Accepted XXX. Received YYY; in original form ZZZ}

\pubyear{2015}

\begin{document}
\label{firstpage}
\pagerange{\pageref{firstpage}--\pageref{lastpage}}
\maketitle

\begin{abstract}
In this work, we employ a soft-sphere discrete element method with a cohesion implementation to model the dynamical process of sub-km-sized cohesive rubble piles under continuous spinup. The dependencies of critical spin periods $T_c$ on several material parameters for oblate rubble piles with different bulk diameters $D$ are explored. Our numerical simulations show that both the increase of interparticle cohesion and particle shape parameter in our model can strengthen the bodies, especially for the smaller ones. In addition, we find there exists some critical diameter $D_{cri,\rho}$ at which the variation trend of $T_c$ with the bulk density $\rho$ reverses. Though a greater static friction coefficient $\mu_S$ can strengthen the body, this effect attains a minimum at a critical diameter $D_{cri,\phi}$ close to $D_{cri,\rho}$. The continuum theory (analytical method) is used for comparison and two equivalent critical diameters are obtained. The numerical results were fitted with the analytical method and the ratio of the interparticle cohesion $c$ to the bulk cohesion $C$ is estimated to be roughly 88.3. We find this ratio keeps constant for different $c$ and $\rho$, while it strongly depends on the friction angle $\phi$. Also, our numerical results further show that the dependency of $T_c$ on $\phi$ is opposite from that predicted by the continuum theory when $D$ < $D_{cri,\phi}$. Finally, we find that the two critical diameters happen to be close to the diameter when the mean normal stress of the body equals zero, which is the separation between the compressive regime and the tensile regime.
\end{abstract}



\section{Introduction}
\begin{figure}
    \includegraphics[width=0.47\textwidth]{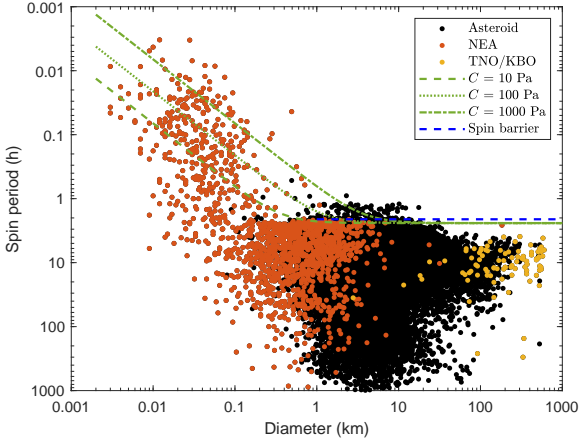}
    \caption{The distribution of spin periods and diameters of asteroids in solar system (the data come from the LCDB database \citep{warner2009asteroid}, updated on 2020 June 26). The green lines with different bulk cohesion $C$ are calculated with the continuum theory by assuming friction angle $\phi$ = 35$^\circ$.}
    \label{fig:T_D_Observation}
\end{figure}

Spin period is one of the important physical parameters for understanding the internal structure of asteroids. Around the turn of the century, it was believed that asteroids were loosely packed, gravity-dominated aggregates, based on the finding at that time that their spin periods were well below a centrifugal barrier around 2.2 hours \citep{harris1996rotation, pravec2000fast}. However, with the accumulation of spin period samples derived from lightcurves and radar observations, a vast number of asteroids are discovered to possess spin periods less than 2.2 hours (see Fig. \ref{fig:T_D_Observation}). In this work, we call these objects super-fast rotators (SFRs), irrespective of their sizes. At the time of this writing, according to the LCDB database, 134 objects larger than 1 km are SFRs (though a few of them may have some considerable uncertainties), 95\% of which are main-belt asteroids (MBAs), while 387 objects smaller than 1 km are SFRs, 94\% of which are near-Earth asteroids (NEAs).

The closeup images and the derived densities from the asteroid missions to (25143) Itokawa, (101955) Bennu and (162173) Ryugu \citep{fujiwara2006rubble, lauretta2019unexpected, watanabe2019hayabusa2}, etc., make it more likely that many asteroids are constructed from collections of aggregates separated by voids, or ``rubble piles", which are considered to be probably formed by asteroid collisions that result in disruption of precursor bodies and re-assembly of fragments \citep{michel2001collisions,johansen2015new}. But we still do not know exactly whether SFRs are rubble piles or not. However, we do have indirect evidences that at least a portion of them can be rubble piles with certain tensile strength among the components.

The low density of km-sized asteroid (29075) 1950 DA measured by Yarkovsky orbital drift and thermal-infrared observations show that it is probably a rubble pile, and the 2.1216 hr spin period requires it to have a minimum cohesion of 44-76 Pa \citep{rozitis2014cohesive} or 75-85 Pa \citep{hirabayashi2014stress}. The disruption event of the active asteroid P/2013 R3 was observed and a rotationally induced structural failure was considered to be the mechanism that triggered the disaggregation \citep{jewitt2014disintegrating}, from which a level of cohesion ranging between 40 Pa and 210 Pa was estimated \citep{hirabayashi2014constraints}. 2008 TC3, an elongated asteroid with the longest length of 6.7 meters rotating in an excited state with a period of rotation of 99.2 s and precession of 97.0 s, entered Earth's atmosphere above northern Sudan on October 7, 2008 \citep{scheirich2010shape, shaddad2010recovery}. The bulk density of 1.8 g/cm$^3$, porosity of $\sim$ 50\%, and heterogeneous composition show that 2008 TC3 was a good rubble pile candidate, with a minimum cohesion level of $\sim$ 25 Pa to hold the components together \citep{sanchez2014strength, borovivcka2015small}, though data on the behavior of 2008 TC3 during the atmospheric entry was too poor to prove it further \citep{borovivcka2015some}.

Data from missions to asteroids (and comets) provide additional direct evidence to estimate the material strength. By analyzing the movement of surface materials on the steep cliffs of comet 67P/Churyumov-Gerasimenko, a tensile strength between 1.5 to 100 Pa was estimated \citep{basilevsky2016estimating}. A unique longitudinal variation in geomorphology was observed on Ryugu, and recent numerical research showed that the smooth surface and sharp equatorial ridge in this area can provide a constraint on cohesion ranging between $\sim$4 Pa and $\sim$10 Pa \citep{hirabayashi2019western}. Based on the constraint of surface stability, \cite{scheeres2019dynamic} estimated the minimum cohesion of (101955) Bennu is at a level of 1 Pa.

Currently, the general consensus indicates that van der Waals force is the main source of cohesive force between constituent regolith grains on asteroid \citep{scheeres2010scaling}. Previous explorations suggest that electrostatic force may play a more important role in some situations, but it is still poorly understood \citep{colwell2005dust, berkebile2012adhesion}. \cite{scheeres2010scaling} analyzed several physical forces that may act on asteroid regolith and found that the van der Waals cohesive force could be as important as the gravity for small asteroids. For NEAs and inner MBAs less than 10 km in diameter, their rotational rates can be accelerated by the Yarkovsky-O'Keefe-Radzievskii-Paddack (YORP) effect that results from the net radiation recoil torques caused by anisotropic re-emitting of photons on irregularly shaped asteroids \citep{rubincam2000radiative, lowry2007direct}. The slow spin-up process caused by the small but continuous force can finally result in surface shedding, deformation, global disintegration, and even the formation of a binary system \citep{walsh2008rotational, walsh2012spin, scheeres2015landslides}. With consideration of both the YORP effect and cohesive force, many efforts have been made to understand the dynamical behaviors of cohesive rubble piles under continuous spinup. A series of numerical investigations have revealed that the heterogeneous internal cohesion distribution, as well as the actual shape, can highly affect the failure mode of a cohesive rubble-pile asteroid due to a quasi-static spinup \citep{hirabayashi2015internal, hirabayashi2015failure, hirabayashi2015failure2, hirabayashi2019rotationally, sanchez2018rotational}, which can be a reason to form the equatorial cavities found on asteroids 2008 EV5 and 2000 DP107 Alpha \citep{tardivel2018equatorial}. Through the use of a soft-sphere discrete element method (SSDEM), \cite{sanchez2016disruption} found that the angle of friction can affect the level of deformation that takes place before disruption, and that higher tensile strength tends to result in a larger amount of fissioned material. \cite{zhang2018rotational} used a different implementation of SSDEM and emphasized that both frictional and cohesive force can greatly influence the critical spin period, while failure mode only shows obvious dependence on the cohesive force. 

Based on the elastic-plastic continuum theory, \cite{holsapple2001equilibrium} and \cite{holsapple2004equilibrium} developed a purely analytical solution to obtain the equilibrium configurations of spinning cohesionless rubble piles by applying the Mohr-Coloumb yield criterion. An equivalent solution using the Drucker-Prager strength model was also developed and applied to cohesive rubble-pile asteroids \citep{holsapple2004equilibrium, holsapple2007spin}. Due to its simplicity, this method has been widely used to constrain the minimum internal cohesion of a triaxial ellipsoid asteroid with given size, spin rate and angle of friction, which can be used to roughly judge whether a fast-rotating asteroid can be rubble pile or not \citep{rozitis2014cohesive, polishook20162, polishook2017fast}.

In this work, we only focus on sub-km-sized cohesive rubble-pile asteroids with diameter ranging between 50 m and 1,000 m. Previous study showed that small bodies may be involved in different disaggregation behaviors due to YORP spinup, and the bodies may enter a ``disaggregation phase" that no binary system can form even for a small level of cohesion \citep{scheeres2018disaggregation}. Rather than discussing the failure mode, we here calculate their critical spin periods ($T_c$) under continuous spinup with a SSDEM method, and investigate the dependencies of $T_c$ on several material parameters. Specially, we aim to explore whether the dependencies can change with the bulk size, in an effort to better understand the dynamical behaviors of small cohesive rubble-pile asteroids at the critically spinning state. In addition, by analyzing the $T_c-D$ curves, we can compare our numerical results with the analytical solution given by \cite{holsapple2007spin}.

\section{Method and model}
\subsection{\texorpdfstring{$pkdgrav$}{Lg} with cohesion}
In this work, we applied a parallel $N$-body tree code, $pkdgrav$, to model the spin-up process of rubble piles and calculate the critical spin periods in a varied parameter space. $pkdgrav$ was originally developed to simulate the large-scale-structure $N$-body problem \citep{stadel2001cosmological} and then was adapted to deal with hard-body collisions \citep{richardson2000direct, richardson2009numerical}. An SSDEM implementation was also added, which allows particles to interpenetrate each other slightly to mimic the deformation at contact \citep{schwartz2012implementation, zhang2017creep}. The code has been widely and successfully used to examine the tidal process \citep{richardson1998tidal, yu2014numerical, demartini2019using, zhang2020tidal}, formation and evolution of binary asteroids \citep{walsh2006binary, walsh2008rotational}, fragment accumulation under self-gravity \citep{michel2013collision, schwartz2018catastrophic}, and the formation of contact binaries \citep{hu2018formation, mckinnon2020solar}. 

Recently, interparticle cohesion caused by van der Waals force was further added in $pkdgrav$, along with an implementation of static, rolling and twisting friction, which makes it possible for us to model the dynamical process of spinning cohesive rubble piles \citep{zhang2018rotational}. In this work, this implementation is employed to simulate the dynamical process of rubble piles subjected to continuous spinup. To better understand the procedure, it is beneficial to briefly revisit the SSDEM modeling with cohesion included. More details about the implementation can be found in \cite{schwartz2012implementation}, \cite{zhang2017creep} and \cite{zhang2018rotational}.

In the implementation, a linear spring-dashpot model is employed to describe the normal contact force ${\bf{F}}_N$ and the tangential stick-slip force ${\bf{F}}_S$ \citep{cundall1979discrete}. In addition, the cohesive force ${\bf{F}}_C$ arises when two particles are in contact. To better capture the quasi-static behavior of a spinning rubble pile, an elastic-plastic spring-dashpot rotational resistance model is applied, in which the resulting torque due to contact can be decomposed into twisting and rolling components, ${\bf{M}}_T$ and ${\bf{M}}_R$.

The idea of a granular bridge is applied in the implementation of the cohesive force, so that the cohesive force between two large boulders is considered as a cumulative effect of the interstitial cohesive grains \citep{sanchez2014strength, sanchez2016disruption}. ${\bf{F}}_C$ is expressed as
\begin{equation}
\label{eq:Fc}
    {{\bf{F}}_C}\;{\rm{ = }}\;c{A_{eff}}{\bf{\hat n}},\;\;\;\;{A_{eff}} = 4{\left( {\beta R} \right)^2},\;R = \frac{{{R_i}{R_j}}}{{{R_i} + {R_j}}}
\end{equation}
where $R_i$ and $R_j$ are the radii of the two neighboring particles, and $c$ is the interparticle cohesion, measured in Pascal. A shape parameter $\beta$, which was first introduced in the contact model of \cite{jiang2013microscopic, jiang2015novel}, is also used in $pkdgrav$ to characterize the size of contact area.

All the dominant equations of the forces and torques are given in Table \ref{tab:force}. If the initial positions, velocities and spin states of constituent particles are known, the fate of a spun-up rubble pile is governed by 13 parameters: normal/tangential spring constants and viscous damping coefficients $k_N$, $k_S$ and $C_N$, $C_S$; rolling/twisting stiffness and damping coefficients $k_R$, $k_T$ and $C_R$, $C_T$; static friction coefficient $\mu_S$ and static rolling/twisting friction coefficients $\mu_R$ and $\mu_T$; interparticle cohesion $c$ and shape parameter $\beta$. Note that
\begin{equation}
    \label{eq:kC}
    \begin{array}{l}
        {k_R} = {k_N}{\left( {\beta R} \right)^2},\;\;\;\;{C_R} = {C_N}{\left( {\beta R} \right)^2}\\
        {k_T} = 2{k_S}{\left( {\beta R} \right)^2},\;\;\;{C_T} = 2{C_S}{\left( {\beta R} \right)^2}
    \end{array}
\end{equation}
Thus we only have 9 free parameters. To further narrow the parameter space, some will be kept constant in our study. For ``gravel"-like material, we have: $\mu_R$ = 1.05, $\mu_T$ = 1.3; $C_N$ and $C_S$ can be obtained by the normal and tangential coefficients of restitution $\epsilon_N$ and $\epsilon_S$, both of which are set to be 0.55 \citep{jiang2015novel}. $k_N$ (as well as the timestep) is determined by ensuring the overlaps not exceed 0.01 of the minimum particles radius; $k_S$ is usually set to (2/7)$k_N$ to keep the normal and tangential
oscillation frequencies equal \citep{schwartz2012implementation, jiang2015novel}. Thus, in addition to the bulk size, a space including four parameters, $c$, $\beta$, $\rho$ and $\mu_S$, will be explored in this work.

\begin{table*}
    \centering
    \caption{Equations of contact forces and torques}
    \label{tab:force}
    \begin{tabular}{lll}
    \toprule
    Forces or torques & Symbols & Equations \\
    \midrule
    Normal force & ${\bf{F}}_N$ & $- {k_N}x{\bf{\hat n}} + {C_N}{{\bf{u}}_n}$ \\
    Tangential force & ${\bf{F}}_S$ & $\min \left( {{k_S}{{\bf{\delta }}_S} + {C_S}{{\bf{u}}_t},\;{\mu _S}\left| {{{\bf{F}}_N}} \right| \cdot {{\bf{\delta }}_S}/\left| {{{\bf{\delta }}_S}} \right|} \right)$ \\
    Cohesive force & ${\bf{F}}_C$ & $c{A_{eff}}{\bf{\hat n}}$ \\
    Rolling torque & ${\bf{M}}_R$ & $\left\{ \begin{array}{l}
        {k_R}{{\bf{\delta }}_R} + {C_R}{\omega _R},\;\;\;|{k_R}{{\bf{\delta }}_R}|\; < \;{M_{R,\max }}\\
        {M_{R,\max }}{{\bf{\delta }}_R}/\left| {{{\bf{\delta }}_R}} \right|,\;\;|{k_R}{{\bf{\delta }}_R}|\; \ge \;{M_{R,\max }}
        \end{array} \right.$\\
    Twisting torque & ${\bf{M}}_T$ & $\left\{ \begin{array}{l}
        {k_T}{{\bf{\delta }}_T} + {C_T}{\omega _T},\;\;\;|{k_T}{{\bf{\delta }}_T}|\; < \;{M_{T,\max }}\\
        {M_{T,\max }}{{\bf{\delta }}_T}/\left| {{{\bf{\delta }}_T}} \right|,\;\;|{k_T}{{\bf{\delta }}_T}|\; \ge \;{M_{T,\max }}
        \end{array} \right.$ \\
    \bottomrule
    \end{tabular}
    \begin{tablenotes}
        \item[1] where $x$ is the overlap, ${M_{R,\max }} = {\mu _R}\beta R\left| {{{\bf{F}}_N}} \right|$ and ${M_{T,\max }} = {\mu _T}\beta R{\mu _S}\left| {{{\bf{F}}_N}} \right|$. Unit vector ${{\bf{\hat n}}}$ is the direction from the center of one particle to its neighbor's. ${\bf{\delta}}_S$ is the sliding displacement from the equilibrium contact point. ${\bf{u}}_n$ and ${\bf{u}}_t$ are the normal and tangential relative velocity, respectively. ${\bf{\delta}}_R$ and ${\bf{\delta}}_T$ are the rolling and twisting angular displacement, respectively. ${\bf{\omega}}_R$ and ${\bf{\omega}}_T$ are the relative rolling and twisting angular velocity, respectively. More detailed expressions can be found in \cite{zhang2018rotational}.
    \end{tablenotes}
\end{table*}

\subsection{Initial conditions}
For a theoretical investigation, we will not focus on any specific asteroids in our simulations. The initial shapes of the test bodies are assumed to be ellipsoids with different sizes characterized by the three semi-axis lengths $a_1$, $a_2$ and $a_3$ $({a_1} \ge {a_2} \ge {a_3})$ and the equivalent diameter $D$ is defined as $D = 2(a_1a_2a_3)^{1/3}$. Ten different diameters ranging from 50 m to 1,000 m with a log-uniform distribution are considered, as given in the second row of Table \ref{tab:sizes}. Oblate spheroids with $\alpha = a_3/a_1 = 0.9$ $(a_1 = a_2)$ are selected as the nominal shape. In view of the fact that $\alpha \simeq 0.92$ and $\alpha \simeq 0.87$ for the top-shaped asteroids Bennu and Ryugu \citep{barnouin2019shape, watanabe2019hayabusa2}, this is a reasonable assumption for fast-spinning rubble-pile asteroids.

\begin{table*}
    \centering
    \caption{The mean bulk diameters $D$ and mean particle diameters $D_P$ of the ten nominal oblate rubble piles (composed of 10,000 particles) used in our simulations.}
    \label{tab:sizes}
    \begin{tabular}{llllllllllll}
    \hline
    NO.     & 1  & 2  & 3 & 4 & 5 & 6 & 7 & 8 & 9 & 10 \\ \hline
    $D$ (m) & 50 & 69.7 & 97.3 & 135.7 & 189.3 & 264.1 & 368.4 & 513.9 & 716.9 & 1000 \\ \hline
    ${D_P}$ (m) & 1.8 & 2.5 & 3.4 & 4.8 & 6.7 & 9.3 & 13.0 & 18.1 & 25.3 & 35.2 \\ \hline
    \end{tabular}
\end{table*}

The structures of the test bodies are composed of a number of spherical particles in contact with a -3-index power-law distribution in size and a ratio of maximum to minimum particle size of 3. The bodies are carved from a much larger parent rubble pile that has been settled down from a randomly distributed particle cloud under self-gravity. The effects of other kinds of particle size distributions (eg., monodisperse particles) and packings (eg., hexagonal closest packing) are not considered in this research, since the polydisperse packing model is a better approximation to real rubble-pile asteroids.

For simplicity, an oblate body with axis ratio defined above is carved from the parent rubble pile, and the ten different-sized rubble piles are obtained by dilating or shrinking this source body. In all the different-sized bodies, we used 10,000 particles to constitute the structure, which is a balance between model precision and computational overhead. The corresponding mean particle diameters $D_P$ are shown in the third row of Table \ref{tab:sizes}. We will quantitatively calculate how much $T_c$ can be changed for different particle arrangements and model precisions in the following section.

The nominal values of the four parameters are: $c$ = 1600 Pa, $\beta$ = 0.5, $\rho$ = 2.4 g/cm$^3$ and $\mu_S$ = 0.5, which corresponds to material with a friction angle of approximately 32.9$^\circ$ \citep{zhang2018rotational}. A varied space of $c$ = 800, 1600 and 3200 Pa, $\beta$ = 0.3, 0.5 and 0.7, $\rho$ = 1.8, 2.4, 3.0 g/cm$^3$, and $\mu_S$ = 0.3, 0.5 and 0.7, will be explored in current work.

\section{Critical spin period}
\subsection{Continuum theory}
Before we continue, it is important to recall the analytical solution derived by \cite{holsapple2007spin}. The volume average shear stresses $({{\bar \sigma }_x},\;{{\bar \sigma }_y},\;{{\bar \sigma }_z})$ over a spinning ellipsoid rubble pile (only a uniformly spinning state with the spin vector aligned with the body z axis is considered in the current work) in the three orthogonal directions are:

\begin{equation}
\label{eq:stress}
\begin{array}{l}
    {{\bar \sigma }_x} = \left( {\rho {\omega ^2} - 2\pi {\rho ^2}G{A_x}} \right)\frac{{a_1^2}}{5}\\
    {{\bar \sigma }_y} = \left( {\rho {\omega ^2} - 2\pi {\rho ^2}G{A_y}} \right)\frac{{a_2^2}}{5}\\
    {{\bar \sigma }_z} = \left( { - 2\pi {\rho ^2}G{A_z}} \right)\frac{{a_3^2}}{5}
    \end{array}
\end{equation}
where $\omega$ is the spin rate, and $G$ is the gravitational constant. The three dimensionless functions $A_x$, $A_y$ and $A_z$ are:
\begin{equation}
\label{eq:Axyz}
\begin{array}{l}
    {A_x} = {\alpha _1}{\alpha _2}\int_0^\infty  {\frac{{{\rm{d}}u}}{{{{\left( {u + 1} \right)}^{3/2}}{{\left( {u + \alpha _1^2} \right)}^{1/2}}{{\left( {u + \alpha _2^2} \right)}^{1/2}}}}} \\
    {A_y} = {\alpha _1}{\alpha _2}\int_0^\infty  {\frac{{{\rm{d}}u}}{{{{\left( {u + 1} \right)}^{1/2}}{{\left( {u + \alpha _1^2} \right)}^{1/2}}{{\left( {u + \alpha _2^2} \right)}^{3/2}}}}} \\
    {A_z} = {\alpha _1}{\alpha _2}\int_0^\infty  {\frac{{{\rm{d}}u}}{{{{\left( {u + 1} \right)}^{1/2}}{{\left( {u + \alpha _1^2} \right)}^{3/2}}{{\left( {u + \alpha _2^2} \right)}^{1/2}}}}}
    \end{array}
\end{equation}
which are related to the axial ratio $\alpha_1=a_3/a_1$ and $\alpha_2=a_2/a_1$ and can be numerically computed. For the oblate shape used in this work, we have ${A_x} = {A_y} = 0.638,\;{A_z} = 0.724$. According to the Drucker-Prager yield criterion, the inequality
\begin{equation}
\label{eq:dpyc}
\begin{array}{l}
    \sqrt {{J_2}}  \le k - 3sp
\end{array}
\end{equation}
should be satisfied to keep the structure intact, in which $J_2$ is the second invariant of the stress deviator tensor
\begin{equation}
    {J_2} = \frac{1}{6}\left[ {{{\left( {{{\bar \sigma }_x} - {{\bar \sigma }_y}} \right)}^2} + {{\left( {{{\bar \sigma }_y} - {{\bar \sigma }_z}} \right)}^2} + {{\left( {{{\bar \sigma }_z} - {{\bar \sigma }_x}} \right)}^2}} \right]
\end{equation}
$p$ is the mean normal stress (the stress is in compression when $p$ < 0 and in tension when $p$ > 0)
\begin{equation}
    \label{eq:mean_normal_stress}
    p = \frac{1}{3}\left( {{{\bar \sigma }_x} + {{\bar \sigma }_y} + {{\bar \sigma }_z}} \right)
\end{equation}
and $k$ and $s$ are defined as \citep{chen2007plasticity}
\begin{equation}
    \begin{array}{l}
    k = \frac{{6C\cos \phi }}{{\sqrt 3 \left( {3 - \sin \phi } \right)}},\;\;s = \frac{{2\sin \phi }}{{\sqrt 3 \left( {3 - \sin \phi } \right)}}
    \end{array}
\end{equation}
where $C$ is the bulk cohesion and $\phi$ is the angle of friction.

If the size, $C$ and $\phi$ of the ellipsoid body are known, we can calculate the critical spin rate $\omega_c$ by solving Eq. \ref{eq:dpyc} with the inequality replaced with an equal sign (we will also call this as an analytical method in the following text). However, the cohesion parameter in our SSDEM method is given by the interparticle cohesion $c$ rather than the bulk cohesion $C$. \cite{zhang2018rotational} showed that the ratio of $c$ to $k$ is $\sim$100 (for $\beta=0.5$). This implies that the nominal interparticle cohesion of 1600 Pa corresponds to a bulk cohesion of about 16 Pa, which is a mild cohesion level according to our current knowledge of asteroids. In the following text, we will also estimate the value of $c/k$ by fitting our numerical results with the analytical method.

\subsection{Determining the critical spin period}
By adding angular momentum continuously to a spinning rubble pile, we can simulate the YORP-induced spin-up process with the SSDEM code, which is done by making the spin rate increase in steps. The whole procedure was described in detailed in \cite{zhang2018rotational}, which is also similar to the simulations of \cite{sanchez2012simulation}. At the beginning, the test body runs freely under its own gravity with a slow starting period $T_0$ for a sufficiently long duration of time $\Delta {t_0}$ to make the constituent particles settle down (phase A). Then it spins up to period $T_1$ in a relatively rapid way within time interval $\Delta {t_1}$ (phase B), and finally slowly spins up to a sufficiently small period $T_2$ within $\Delta {t_2}$ (phase C), so that the critical spin period $T_c$ lies between $T_1$ and $T_2$. Using Eq. (\ref{eq:dpyc}), we can obtain a rough estimation on the critical spin period ${T_c}'$ by assuming $c/k = 100$. Then we used $T_1$ = min$(T_0, 2{T_c}')$ and $T_2 = {T_c}'/2$ in our simulations. In practice, we set $T_0 = 6$ h for all the cases, which is enough for the bodies to settle down. Our numerical experiments show that the obtained $T_c$ is located well within $T_1$ and $T_2$ for any set of parameters we considered.

\begin{figure}
    \includegraphics[width=0.47\textwidth]{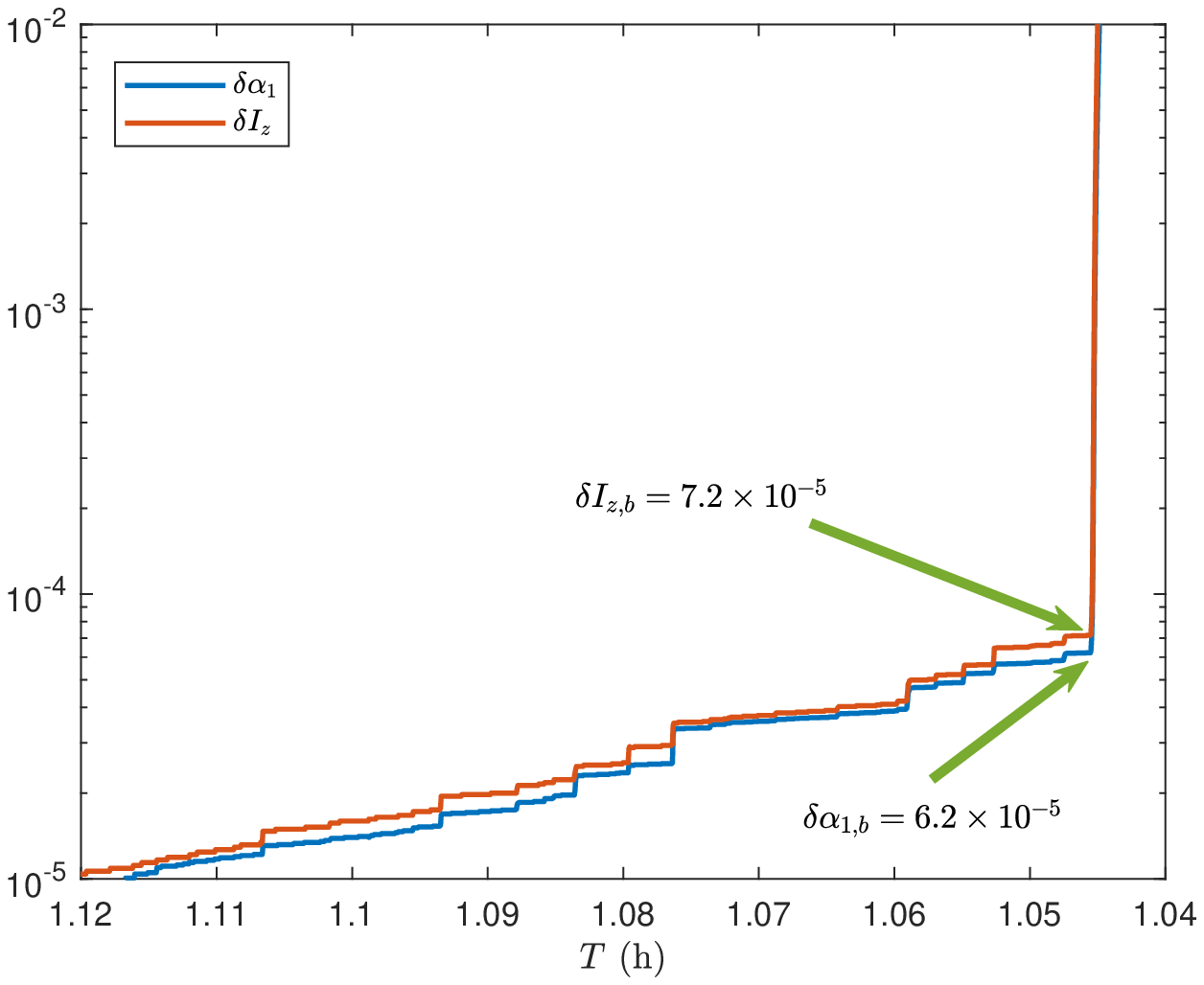}
    \caption{The relative changes of axis ratio ($\delta {\alpha _1} = \frac{{\left| {{\alpha _1}\left( t \right) - {\alpha _1}\left( {{t_0}} \right)} \right|}}{{{\alpha _1}\left( {{t_0}} \right)}}$) and moment of inertia ($\delta {I_z} = \frac{{\left| {{I_z}\left( t \right) - {I_z}\left( {{t_0}} \right)} \right|}}{{{I_z}\left( {{t_0}} \right)}}$) over spin period ($T$) during the spinup (absolute values are taken). Parameters in this case are: $D$ = 264.1 m, $\rho$ = 2.4 g/cm$^3$, $\mu_S$ = 0.5, $c$ = 1600 Pa and $\beta$ = 0.5. The global failure occurs when $\delta \alpha_1$ = $6.2\times10^{-5}$ or $\delta I_z$ = $7.2\times10^{-5}$.}
    \label{fig:diff-T0}
\end{figure}

The critical spin period $T_c$ is determined at the moment when a global failure occurs, which is obtained when the ratio of axis length $\alpha_1$ or moment of inertia $I_z$ changes by some amount. Taking the oblate case ($D$ = 264.1 m) with the nominal parameters as an example, the relative change of $\alpha_1$ and $I_z$ over spin period during the spinup are shown in Fig. \ref{fig:diff-T0}, in which only the segment with $T$ < 1.12 h is illustrated. The curves show that both of them gradually increase as $T$ continuously decreases, and the zigzags on the curves demonstrate that minor interparticle adjustments occur as the spin rate increases. From the figure, we see that global failure is triggered when the relative changes of $\alpha_1$ and $I_z$ at breakup, $\delta{\alpha_{1,b}}$ and $\delta{I_{z,b}}$, reach $6.2\times10^{-5}$ and $7.2\times10^{-5}$, respectively. Then we have $T_c$ = 1.04548 h and 1.04549 h at $\delta{\alpha_{1,b}}$ and $\delta{I_{z,b}}$, respectively, which gives an insignificant difference of less than 0.001\% in $T_c$. In most situations, these two criteria are equivalent to each other, but a series of numerical experiments show that the $\delta{I_z}$ criterion is more robust and gives a smoother $T_c - D$ curve, since some surface particles may be located in unstable areas, which may cause a relatively greater variation of $\alpha_1$ earlier than the more definite global failure and result in a higher $T_c$. So we will adopt $\delta{I_z}$ as the criterion to determine $T_c$ in this work.

For phases A and B, we only need to choose $\Delta{t_0}$ and $\Delta{t_1}$ so that the particles settle down at the end of phase A and still hold the initial shape at the end of phase B. For phase C, however, if the duration is too short, the particle aggregates will not have enough time to relax and adjust during the spinup, which usually results in an unphysical $T_c$ smaller than the real value. But a larger $\Delta{t_2}$ always results in a higher computation burden. We address this by testing several different $\Delta{t_2}$, and plotting a curve of $T_c$ with respect to $\Delta{t_2}$. Then we can see that as $\Delta{t_2}$ increases, $T_c$ tends to get stable for some $\Delta{t_2}$, which can be served as an approach to get a relatively precise $T_c$ while keeping a relatively low computation effort. In practice, we found that $\Delta{t_2} = 2 $ days is a good choice to ensure that the relative error of $T_c$ is less than 1\%.

\begin{figure}
    \includegraphics[width=0.47\textwidth]{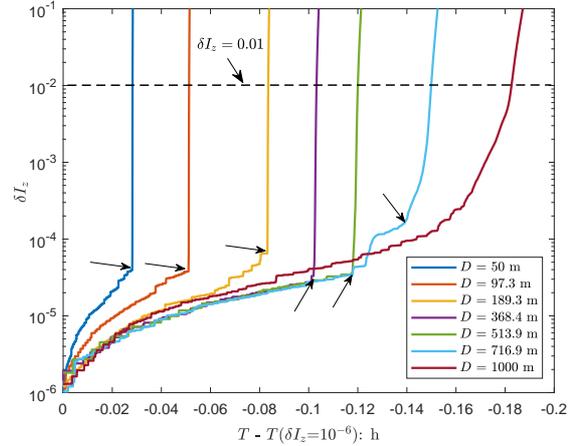}
    \caption{The relative changes of moment of inertia over spin period during the spinup. The spin period is expressed as a deviation from $T$($\delta I_z$=10$^{-6}$). Parameters are the same as those taken in Fig. \ref{fig:diff-T0}, except that seven different diameters are considered. The positions of global failures for $D$ < 1000 m are labeled by arrows.}
    \label{fig:Iz_diff_D}
\end{figure}

According to Fig. \ref{fig:diff-T0}, once global failure starts, whether we choose $\delta I_z$ = 10$^{-4}$, 10$^{-3}$ or 10$^{-2}$ as the criterion to obtain $T_c$ is unimportant (of course it should at least exceed $7.2\times10^{-5}$). This can be also warranted for other diameters, as shown for cases with $D$ < 1000 m in Fig. \ref{fig:Iz_diff_D}, in which global failure can be easily determined with eyes according to the moment when $\delta I_z$ rise sharply at inflection points, which have been labeled in the figure. For simplicity, we will adopt $\delta I_z$ = 10$^{-2}$ as the criterion to determine the critical spin period. Such a practice is justified for these cases. However, situations can be more complicated for $D$ = 1,000 m, for which the body experiences deformation rather than a violent distruction and the resulting $\delta I_z-T$ curve rises more gently. Nevertheless, the deformation has caused significant deviation from the original shape, and the period at the moment ($\delta I_z$ = 10$^{-2}$) is still taken as the critical spin period. This will inevitably lead to a higher uncertainty in $T_c$. However, this uncertainty is insignificant (for the case $D$ = 1000 m in Fig. \ref{fig:Iz_diff_D}, the difference of $T$ within $\delta I_z$ = 10$^{-2}$ and 10$^{-4}$ is only about 1.3\%) and will not affect the main conclusions of this work.

\subsection{Uncertainty of \texorpdfstring{$T_c$}{Lg} caused by particle arrangement and resolution}
Apart from the material parameters $c$, $\beta$, $\rho$ and $\mu_S$, the unknown internal structure of rubble piles can result in some uncertainty in the critical spin period. Specifically, the arrangement of the constituent particles and the rubble-pile model resolution (characterized by the particle number $N$) can affect the contact network and eventually impact the critical spin period. It is important for us to quantitatively estimate how the critical spin period can vary for different particle arrangements and different particle numbers.

\begin{figure*}
    \includegraphics[width=1.0\textwidth]{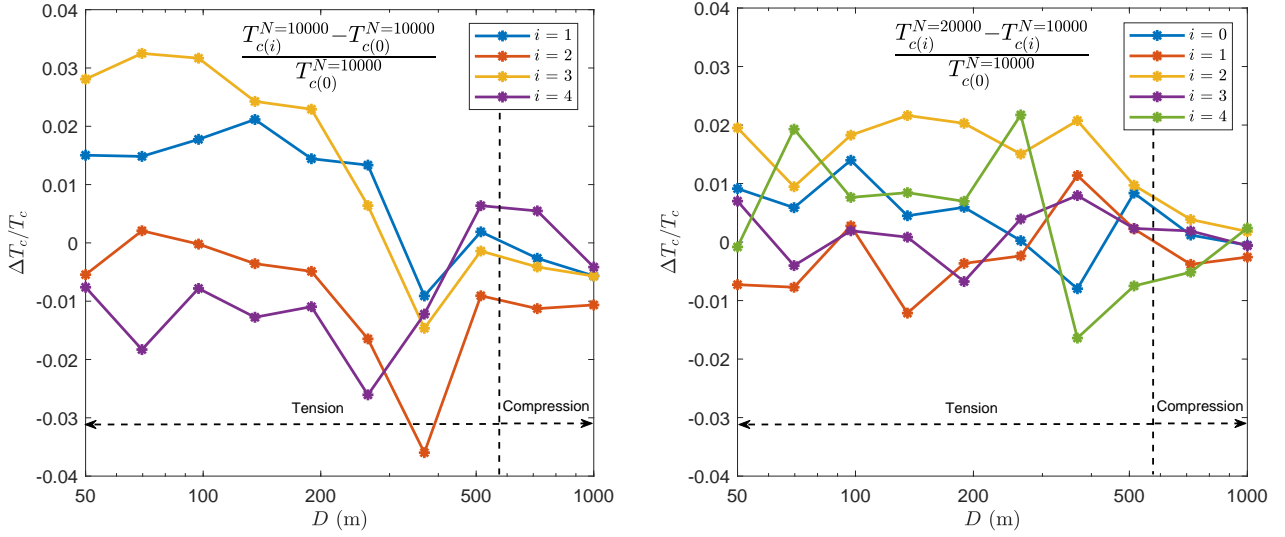}
    \caption{The relative differences of critical spin periods with different particle arrangements (left) and different model resolutions (right). The nominal parameters $c$ = 1600 Pa, $\beta$ = 0.5, $\rho$ = 2.4 g/cm$^3$ and $\mu_s$ = 0.5 are used.}
    \label{fig:Tc-diff_structure}
\end{figure*}

As mentioned above, the same arrangement is applied to all the test rubble-pile models. We can change the arrangement by shifting the carving center, or altering the orientation of the parent rubble pile relative to the inertial reference frame randomly. Here we have considered five different arrangements with $N$ = 10,000 and the resulting critical spin periods, $T_{c(i)}^{N = 10000}$ ($i$ = 0,$\ldots$,4, where $i$ = 0 corresponds to the nominal arrangement used in this work), are calculated with the above criterion. The relative differences between $T_{c(i)}^{N = 10000}$ ($i$ = 1,$\ldots$,4) and the nominal value $T_{c(0)}^{N = 10000}$ are shown in the left panel of Fig. \ref{fig:Tc-diff_structure}, from which we can see that the average variation of $T_c$ resulting from the uncertainty of particle arrangement is about 2\%, with a maximum of about 4\%.

Due to the limitation of computational resources, the structure of a rubble-pile asteroid contains far more particles than we can model. The nominal value of particle number $N$ is taken as 10,000 in this work, but it is necessary for us to evaluate the difference when increasing the number. Here we have also calculated the critical spin periods for $N$ = 20,000. Their differences in $T_c$ for varied diameters are presented in the right panel of Fig. \ref{fig:Tc-diff_structure}, in which the five different arrangements are also considered. We can see that the average difference is only about 1\% and the maximum is about 2\%.

\begin{figure}
    \includegraphics[width=0.47\textwidth]{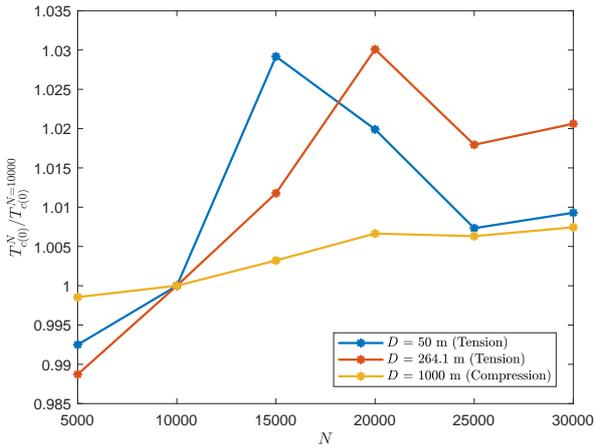}
    \caption{The critical spin periods (divided by $T_{c(0)}^{N=10000}$) of $D$ = 50 m, 264.1 m and 1,000 m for different particle numbers from 5,000 to 30,0000. The nominal material parameters and the nominal particle arrangement are used.}
    \label{fig:Tc_diffN_diffD}
\end{figure}

For a broader range of $N$ from 5,000 to 30,0000 with an increment of 5,000, the results of $T_c$ (normalized by $T_{c(0)}^{N=10000}$) for $D$ = 50 m, 264.1 m and 1,000 m, as well as the nominal arrangement and the other nominal parameters, are plotted in Fig. \ref{fig:Tc_diffN_diffD}. The curves do not monotonically change when $N$ increases. However, it shows that $T_c$ increases as $N$ changes from 5,000 to 15,000, but finally converges when $N$ > 20,000. The difference of $T_c$ between $N$ = 10,000 and 30,000 is less than $\sim$2\%. These results demonstrate that our model is robust and we can safely use $N$ = 10,000 to calculate $T_c$, with an uncertainty of $\sim$2\%.

An interesting feature of Fig. \ref{fig:Tc-diff_structure} is that the uncertainties of $T_c$ in both panels are generally larger in the tensile regime than in the compressive regime (the two regimes will be clarified in detail in Section 4.7). From Fig. \ref{fig:Tc_diffN_diffD}, we also find that the difference of $T_c$ between $N$ = 10,000 and 30,000 for $D$ = 1,000 (in compression) is evidently smaller than the other two (in tension). These imply that the body strength may be slightly more sensitive to the unknown internal structure in the tensile regime.

Of course, the uncertainties of $T_c$ analyzed above are not present in the continuum theory. However, it is a natural characteristic of the SSDEM model, which physically reflects the inhomogeneous internal structure of rubble piles. Their influence to our conclusions will be discussed in the following section.

\section{Results}
\subsection{Effect of \texorpdfstring{$D$}{Lg}}
The diameter $D$ is one of the most important parameters that can affect the terminal state of a rubble pile subjected to spinup. The $\delta I_z - T$ curves for varied diameters shown in Fig. \ref{fig:Iz_diff_D} clearly indicate that more violent failures occur for smaller bodies. Using the $\delta {I_z}$ = 0.01 criterion, the critical spin periods with varied parameters are calculated with the SSDEM simulations and presented in Table \ref{tab:Tcs-all}. As expected, the results indicate that $T_c$ always decreases as $D$ decreases, but the dependences are different for different parameters. The detailed analysis is carried out in the following sections.

\begin{table*}
    \centering
    \caption{Critical spin periods (measured in hours) of the nominal oblate rubble piles ($T_{c_i}$, $i$ = 1-10, ordered with the ten diameters given in Table \ref{tab:sizes}) obtained from numerical simulations with different parameters.}
    \label{tab:Tcs-all}
    \begin{tabular}{ccccccccccccccc}
        \hline
        NO. & $c$ & $\beta$ & $\rho$ & $\mu_S$ & $T_{c_1}$ & $T_{c_2}$ & $T_{c_3}$ & $T_{c_4}$ & $T_{c_5}$ & $T_{c_6}$ & $T_{c_7}$ & $T_{c_8}$ & $T_{c_9}$ & $T_{c_{10}}$ \\
        \hline
        (1) & 800 & 0.5 & 2.4 & 0.5 & 0.314  & 0.438  & 0.597  & 0.808  & 1.058  & 1.355  & 1.670  & 1.912  & 2.089  & 2.234  \\
        (2) & 1600 & 0.5 & 2.4 & 0.5 & 0.225  & 0.309  & 0.430  & 0.591  & 0.797  & 1.045  & 1.366  & 1.654  & 1.905  & 2.080  \\
        (3) & 3200 & 0.5 & 2.4 & 0.5 & 0.160  & 0.221  & 0.307  & 0.424  & 0.584  & 0.791  & 1.049  & 1.327  & 1.652  & 1.893  \\
        (4) & 1600 & 0.3 & 2.4 & 0.5 & 0.378  & 0.524  & 0.709  & 0.949  & 1.249  & 1.520  & 1.805  & 2.012  & 2.184  & 2.322  \\
        (5) & 1600 & 0.7 & 2.4 & 0.5 & 0.158  & 0.220  & 0.308  & 0.425  & 0.586  & 0.787  & 1.047  & 1.343  & 1.639  & 1.898  \\
        (6) & 1600 & 0.5 & 1.8 & 0.5 & 0.194  & 0.270  & 0.373  & 0.519  & 0.710  & 0.958  & 1.254  & 1.597  & 1.963  & 2.230  \\
        (7) & 1600 & 0.5 & 3.0 & 0.5 & 0.249  & 0.346  & 0.475  & 0.648  & 0.860  & 1.121  & 1.401  & 1.639  & 1.812  & 1.955  \\
        (8) & 1600 & 0.5 & 2.4 & 0.3 & 0.244  & 0.339  & 0.466  & 0.640  & 0.857  & 1.116  & 1.401  & 1.690  & 1.958  & 2.223  \\
        (9) & 1600 & 0.5 & 2.4 & 0.7 & 0.219  & 0.304  & 0.417  & 0.576  & 0.796  & 1.060  & 1.332  & 1.655  & 1.910  & 2.063  \\
        \hline
    \end{tabular}
\end{table*}

The critical spin period can be also calculated by using the analytical solution Eq. \ref{eq:dpyc}. Note that the gravity becomes less important as $D$ decreases. For the case $D$ = 50 m (see case (2) of Table \ref{tab:Tcs-all}), we find
$$
\frac{{2\pi {\rho ^2}G{A_x}}}{{\rho {\omega ^2}}} \approx \frac{{{F_{G}}}}{{{F_{Ct}}}} \approx 0.01
$$
at the critically spinning state ($F_G$ is the surface gravity and $F_{Ct}$ is the centrifugal force at the equatorial surface), which means ${F_G} \ll {F_{Ct}}$. Thus we can remove the gravity term from Eq. \ref{eq:stress} in this situation, and the stress component is simplified as
\begin{equation}
    \label{eq:stress2}
    {\bar \sigma _x} = \frac{{a_1^2}}{5}\rho {\omega ^2},\;{\bar \sigma _y} = \frac{{a_2^2}}{5}\rho {\omega ^2},\;{\bar \sigma _z} = 0
\end{equation}
and then we have the following simpler expression of $T_c$ for a general oblate body
\begin{equation}
    \label{eq:Tc_simple}
    {T_c} = 2\pi \sqrt {\frac{{\rho a_1^2\left( {\frac{1}{{\sqrt 3 }} + 2s} \right)}}{{5k}}}
\end{equation}
which can be simplified further for the nominal oblate body ($a_3/a_1$ = 0.9 and $\phi$ = 32.9$^\circ$)
\begin{equation}
    \label{eq:Tc_simple_nominal}
    {T_c} = 1.4\sqrt {\frac{{\rho {D^2}}}{C}}
\end{equation}
The results of case (2) in Table \ref{tab:Tcs-all} show that
$$
\frac{{{T_c}\left( {D = 50\;{\rm{m}}} \right)}}{{{T_c}\left( {D = 69.7\;{\rm{m}}} \right)}} = \frac{{0.225\;{\rm{h}}}}{{0.309\;{\rm{h}}}} \approx \frac{{50}}{{69.7}}
$$
which can be also predicted with the relationship
$$
T_c \sim D
$$
given by Eq. \ref{eq:Tc_simple} if the gravity is ignored.

Take $\epsilon$ as a small number, say $\epsilon$ = 0.02. When the gravity can be ignored, it means
$$
{F_G} \le \epsilon {F_{Ct}}
$$
In this situation, Eq. \ref{eq:Tc_simple_nominal} can be applied, and we can evaluate the diameter when the gravity can be ignored
\begin{equation}
    \label{eq:D_gravity_ignorable}
    D \le \left( {85\;{\rm{m}}} \right)\sqrt {\varepsilon /(0.02)} \frac{{\sqrt {C/\left( {20\;{\rm{Pa}}} \right)} }}{{\rho /\left( {2\;{\rm{g/c}}{{\rm{m}}^3}} \right)}}
\end{equation}

\subsection{Effect of \texorpdfstring{$c$}{Lg} and \texorpdfstring{$\beta$}{Lg}}
\begin{figure*}
\includegraphics[width=1.0\textwidth]{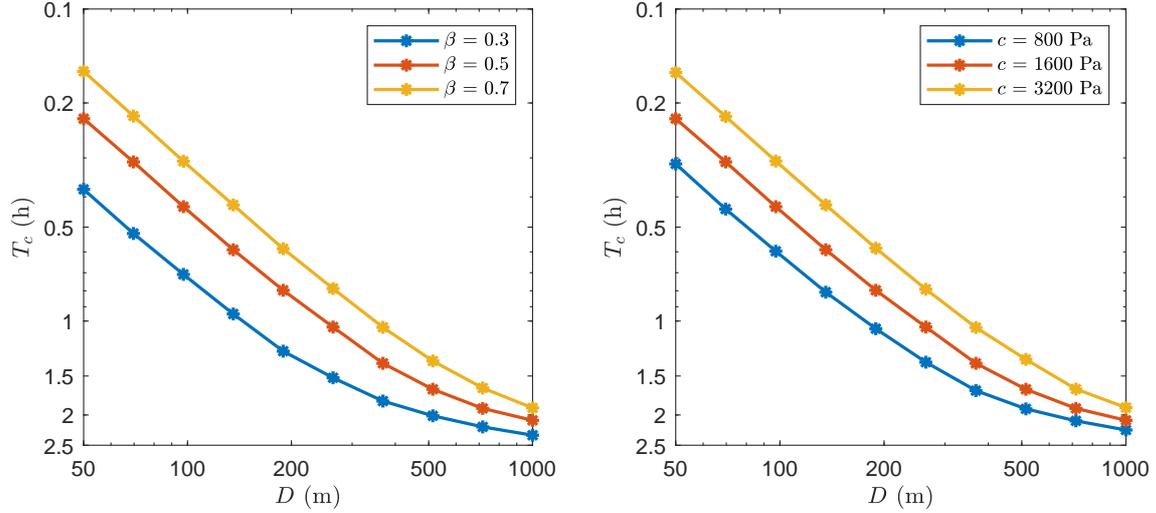}
\caption{Numerical results of $T_c$ with respect to $D$ ($\rho$ = 2.4 g/cm$^3$ and $\mu_S$ = 0.5) for different $\beta$ (left, $c$ = 1600 Pa) and $c$ (right, $\beta$ = 0.5). The results are marked as asterisks and points are connected by straight lines.}
\label{fig:Tc-D-beta-c}
\end{figure*}

Eq. \ref{eq:Fc} indicates that both the interparticle cohesion $c$ and the particle shape parameter $\beta$ can affect the cohesive force, which can strengthen the interparticle bond as they increase. The calculated $T_c$ from SSDEM simulations of $c$ = 800 Pa, 1600 Pa, 3200 Pa and $\beta$ = 0.3, 0.5, 0.7 (as well as with other nominal parameters) are shown in cases (1)-(5) of Table \ref{tab:Tcs-all} and plotted in Fig. \ref{fig:Tc-D-beta-c}.

As expected, $T_c$ decreases as $\beta$ or $c$ increases, with 58.2\% and 18.3\% reduction of $T_c$ when $\beta$ increases from 0.3 to 0.7, and 49.0\% and 15.3\% reduction when $c$ increases from 800 Pa to 3200 Pa, for $D$ = 50 m and $D$ = 1000 m, respectively. The trend is clear that both $\beta$ and $c$ play a more important role in determining $T_c$ for smaller rubble piles (recall $\beta$ determines the relative contact area), which is consistent with the fact that the Bond number (the ratio of cohesive force to gravity of a particle on the surface, as defined in \cite{scheeres2010scaling}) increases as $D$ decreases, which in turn enhances the importance of cohesion for smaller rubble piles. This indicates that it is important to necessarily take account of the combined contributions of $c$ and $\beta$ when modeling the spinup of small cohesive rubble piles with this model.

A simple relationship between $T_c$ and $\beta\sqrt{c}$ can be noted from our results if the diameter satisfies Eq. \ref{eq:D_gravity_ignorable}. For example, based on the results of cases (1), (3) and (4), (5) in Table \ref{tab:Tcs-all} at $D$ = 50 m, we can see that
$$
\frac{{{T_c}\left( {c = 3200\;{\rm{Pa}}} \right)}}{{{T_c}\left( {c = 800\;{\rm{Pa}}} \right)}} = \frac{{0.160\;{\rm{h}}}}{{0.314\;{\rm{h}}}} \approx \sqrt {\frac{{800}}{{3200}}}
$$
and
$$
\frac{{{T_c}\left( {\beta  = 0.7} \right)}}{{{T_c}\left( {\beta  = 0.3} \right)}} = \frac{{0.158\;{\rm{h}}}}{{0.378\;{\rm{h}}}} \approx \frac{{0.3}}{{0.7}}
$$
That is, the relationship
$$
{T_c} \sim \frac{1}{{\beta \sqrt c }}
$$
holds when the diameter is very small. Actually, in this situation, we have
$$
{T_c} \sim \frac{1}{{\sqrt C }}
$$
according to Eq. \ref{eq:Tc_simple_nominal}, which is consistent with our numerical results (we will demonstrate that the interparticle cohesion is proportional to the bulk cohesion in Section 4.5), except that our model also considers the contribution of the contact area.

\subsection{Effect of \texorpdfstring{$\rho$}{Lg}}
\begin{figure*}
    \includegraphics[width=1.0\textwidth]{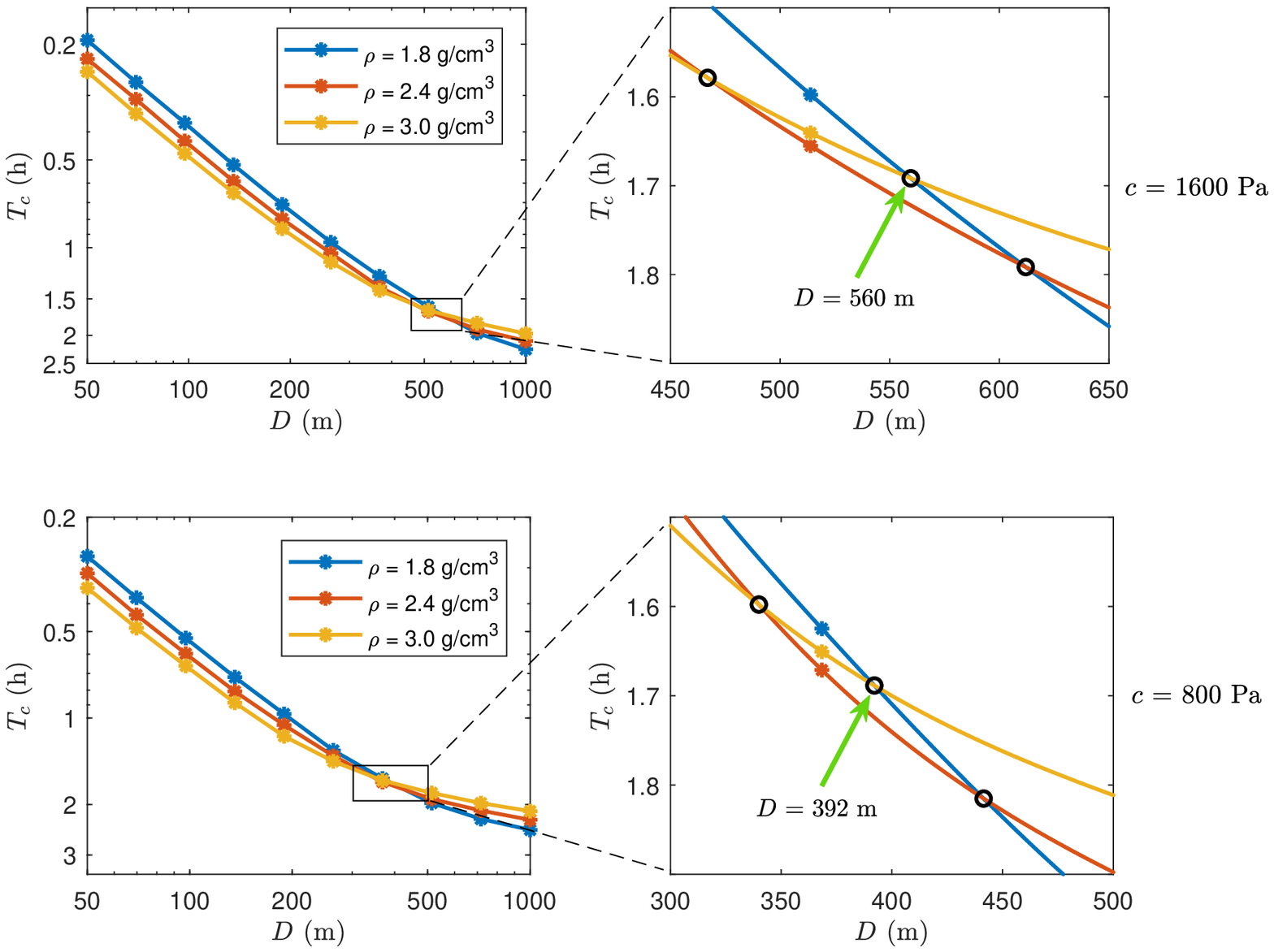}
    \caption{Numerical results of $T_c$ with respect to $D$ (marked as asterisks) for cases of $\beta$ = 0.5, $\mu_S$ = 0.5, and different bulk densities ($c =$ 1600 Pa for the top panels and 800 Pa for the bottom panels). The intersection areas are enlarged and shown in the right panels. Cubic spline interpolations are applied to the points and smooth lines are obtained to find the intersections, which are marked as black circles in the right panels.}
    \label{fig:fig-density-num-oblate-800Pa}
\end{figure*}

The $T_c - D$ curves with $\rho$ = 1.8, 2.4 and 3.0 g/cm$^3$ and the other nominal parameters are shown in the top panels of Fig. \ref{fig:fig-density-num-oblate-800Pa}. As a comparison, results of cases with $c$ = 800 Pa are also given in the bottom panels. A remarkable observation is that $T_c$ does not show a monotonous variation with $\rho$ for different diameters, but there exists some critical diameter $D_{cri,\rho}^N$ (the notation $N$ means it is given by numerical simulations, to differ from the analagous quantity $D_{cri,\rho}^A$ from the analytical theory; these are also collectively referred to as $D_{cri,\rho}$ if not specified and the notation is also applied to $D_{cri,\phi}^N$, $D_{cri,\phi}^A$ and $D_{cri,\phi}$ in the following text) at which the trend in variation of $T_c-\rho$ reverses, which can be seen from both the $c$ = 1600 Pa and 800 Pa cases. That is, $T_c$ decreases as $\rho$ increases when the gravity is more important ($D$ > $D_{cri,\rho}^N$), while $T_c$ increases as $\rho$ increases when $D$ < $D_{cri,\rho}^N$. This can be explained by noting that higher density not only results in stronger gravity but also stronger centrifugal force at the critical limit, but the former strengthens the body while the latter makes the body easier to break up, and the two effects can finally balance with each other at some critical diameter.

However, the three curves with different $\rho$ intersect each other in pairs rather than in a single point. We have enlarged these areas and shown them in the right panels of Fig. \ref{fig:fig-density-num-oblate-800Pa}. This implies that the critical diameter $D_{cri,\rho}^N$ is a function of $\rho$. It is seen that the intersections shift left as $c$ decreases, as revealed by comparing the results of $c$ = 1600 Pa and $c$ = 800 Pa, since a lower cohesion corresponds to a smaller critical spin rate and thus a gentler centrifugal effect, which enhances the importance of gravity and eventually the critical diameter decreases.

Strictly speaking, $D_{cri,\rho}^N$ can be found by solving
$$
    \frac{{\partial {T_c}}}{{\partial \rho }} = 0
$$
which can be approximately expressed as the central difference scheme
\begin{equation}
    \label{eq:chafenfangcheng}
    \frac{{{T_c}\left( {\rho  + \Delta \rho } \right) - {T_c}\left( {\rho  - \Delta \rho } \right)}}{{2\Delta \rho }} = 0
\end{equation}
where ${\Delta \rho}$ is a small density interval. Thus $D_{cri,\rho}^N$ can be approximately determined according to the intersection between the curves of ${{T_c}\left({\rho+\Delta\rho}\right)}$ and ${{T_c}\left({\rho-\Delta\rho}\right)}$ with respect to $D$. Using the cubic spline interpolation, we are able to get a smooth $T_c-D$ curve, and $D_{cri,\rho=2.4\rm{g/cm^3}}^N$ can be calculated with the $T_c$ ($\rho$=1.8 g/cm$^3$) and $T_c$ ($\rho$=3.0 g/cm$^3$) curves, which gives 560 m and 392 m for $c$ = 1600 Pa and $c$ = 800 Pa, respectively, as shown in the right panels of Fig. \ref{fig:fig-density-num-oblate-800Pa}.

\begin{figure*}
    \includegraphics[width=1.0\textwidth]{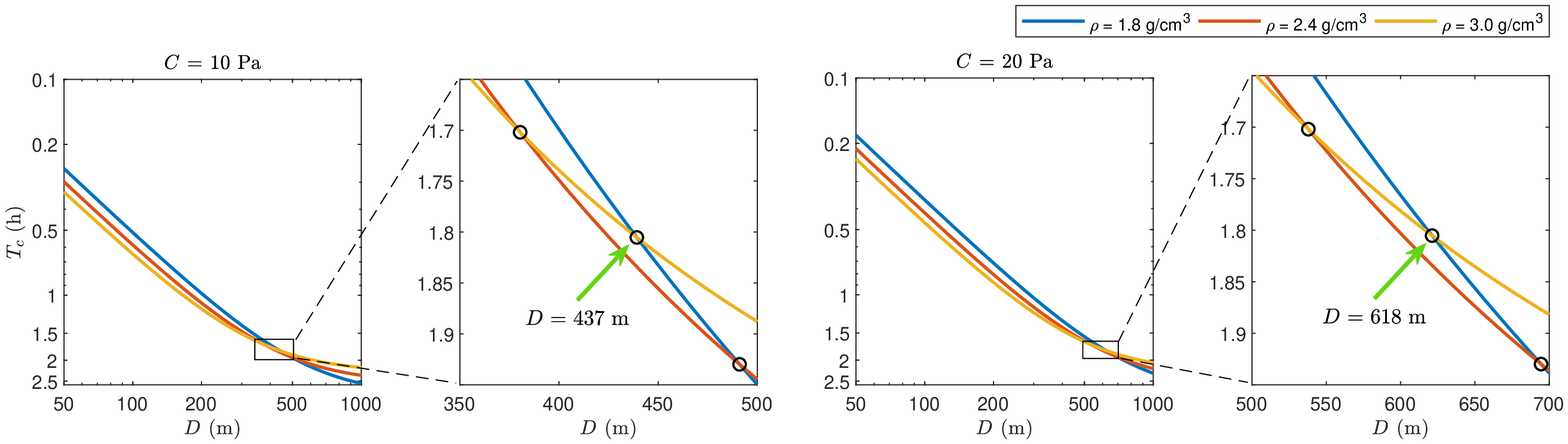}
    \caption{The variations of $T_c$ with respect to $D$ for $C$ = 10 Pa and 20 Pa obtained using the continuum theory ($\phi$ = 32.9$^\circ$). Results are given for $\rho$ = 1.8, 2.4 and 3.0 g/cm$^3$. The intersection areas are enlarged and shown in the 1st and 3rd panels.}
    \label{fig:fig_density_Hol_diffC3}
\end{figure*}

In fact, we find that the opposite $T_c-\rho$ trend at $D<D_{cri,\rho}^N$ and $D>D_{cri,\rho}^N$ can be also observed by applying the continuum theory. The $T_c-D$ results for $\rho$ = 1.8, 2.4 and 3.0 g/cm$^3$ and $C$ = 10 Pa and 20 Pa are shown in Fig. \ref{fig:fig_density_Hol_diffC3}. The intersection area moves from left to right when $C$ increases and the enlarged panels show that the curves do not intersect in a single point but in pairs for each case, both of which are consistent with our numerical findings. With the similar method, the critical diameters $D_{cri,\rho=2.4\rm{g/cm^3}}^A$ are estimated to be 437 m and 618 m, respectively, as labeled in the 1st and 3rd panels.

The results of $T_c$ in Table \ref{tab:Tcs-all} for $D$ = 50 m and $\rho$ = 1.8, 2.4 and 3.0 g/cm$^3$ show that
$$
\frac{{{T_c}\left( {\rho  = 1.8\;{\rm{g/c}}{{\rm{m}}^3}} \right)}}{{{T_c}\left( {\rho  = 2.4\;{\rm{g/c}}{{\rm{m}}^3}} \right)}} = \frac{{0.194\;{\rm{h}}}}{{0.225\;{\rm{h}}}} \approx \sqrt {\frac{{1.8}}{{2.4}}}
$$
and
$$
\frac{{{T_c}\left( {\rho  = 3.0\;{\rm{g/c}}{{\rm{m}}^3}} \right)}}{{{T_c}\left( {\rho  = 2.4\;{\rm{g/c}}{{\rm{m}}^3}} \right)}} = \frac{{0.249\;{\rm{h}}}}{{0.225\;{\rm{h}}}} \approx \sqrt {\frac{{3.0}}{{2.4}}}
$$
which implies
$$
{T_c} \sim \sqrt \rho
$$
holds at small diameter when the gravity is ignorable. We see that this can be also predicted by Eq. \ref{eq:Tc_simple_nominal}.

\subsection{Effect of \texorpdfstring{$\mu_S$}{Lg}}
$\mu_S$ is an important parameter that affects the friction resistance between particles. A greater $\mu_S$ requires more effort to disturb the structure and results in a higher friction angle, as shown in \cite{zhang2018rotational}, in which $\mu_S$ = 0.3, 0.5 and 0.7 ($\beta$ = 0.5) correspond to $\phi$ = 30.6$^\circ$, 32.9$^\circ$ and 34.4$^\circ$, as determined from spinup tests.

\begin{figure*}
    \includegraphics[width=1.0\textwidth]{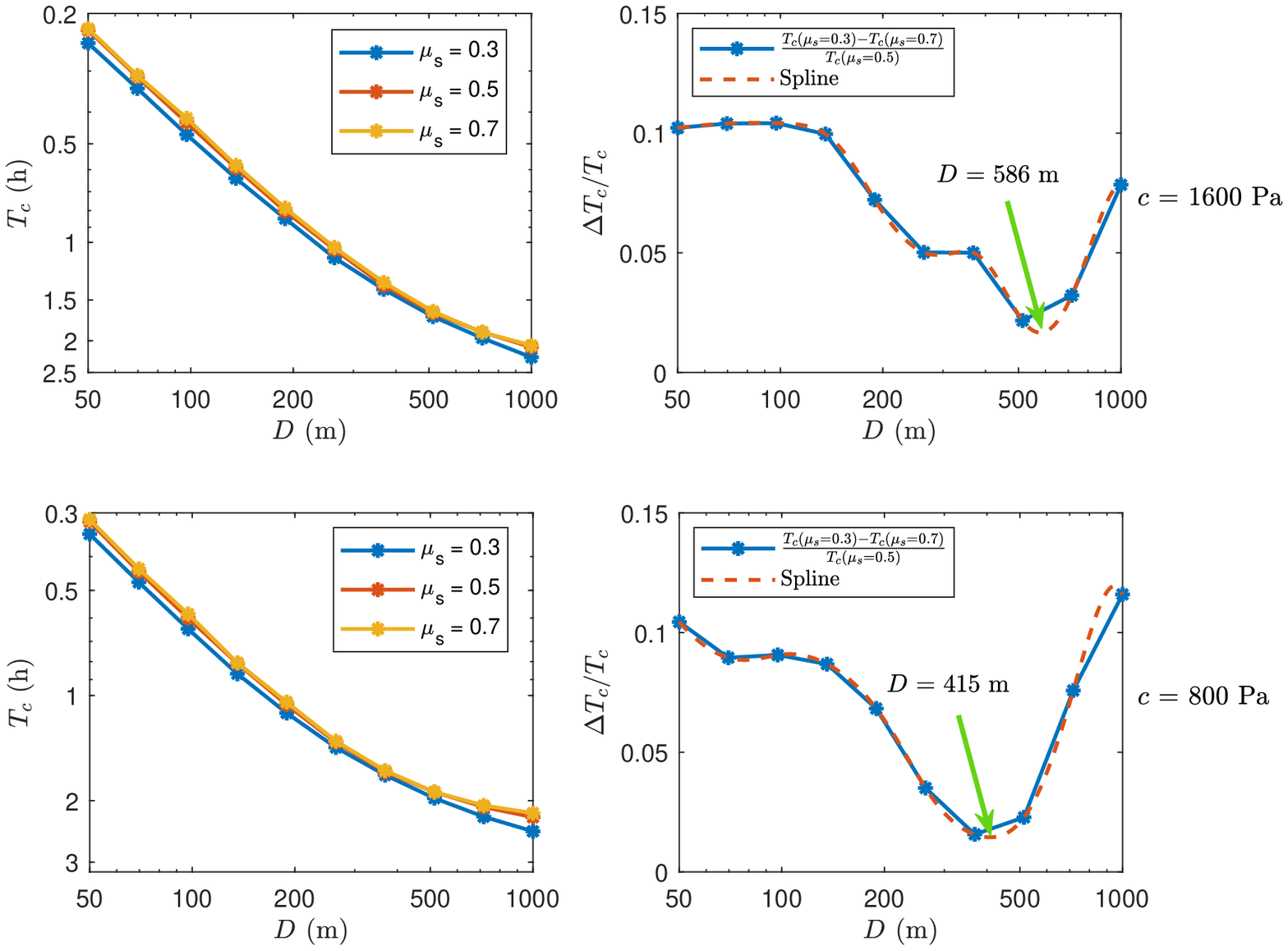}
    \caption{Numerical results of $T_c-D$ (marked as asterisks) for different $\mu_S$ ($c$ = 1600 Pa for the top panels and 800 Pa for the bottom panels). The other parameters are $\rho$ = 2.4 g/cm$^3$ and $\beta$ = 0.5. The relative differences of $T_c$ between $\mu_S$ = 0.3 and $\mu_S$ = 0.7 are shown in the right panels, in which the dashed red curves are given by the cubic spline interpolation and are used to find the critical diameters $D_{cri, \mu_S}$, as labeled by the green arrows.}
    \label{fig:fig-mus-num-oblate-800Pa}
\end{figure*}

For cases of $\mu_S$ = 0.3, 0.5 and 0.7 ($\beta$ = 0.5, $\rho$ = 2.4 g/cm$^3$), the variations of $T_c$ with respect to $D$ are given in the top ($c$ = 1600 Pa) and bottom panels ($c$ = 800 Pa) of Fig. \ref{fig:fig-mus-num-oblate-800Pa}, from which we notice that $T_c$ always decreases as $\mu_S$ increases, consistent with our expectation.

However, according to the difference between $T_c (\mu_S = 0.3)$ and $T_c (\mu_S = 0.7)$ shown in the right panels of Fig. \ref{fig:fig-mus-num-oblate-800Pa}, we find that $\mu_S$ has a minimum effect on $T_c$ at a critical diameter. We know that the friction angle depends strongly on $\mu_S$; this critical diameter is accordingly denoted by $D_{cri,\phi}^N$. For $c = 1600$ Pa and $800$ Pa, we have $D_{cri,\phi}^N$ = 586 m and 415 m, respectively, both of which are close to the corresponding $D_{cri,\rho}^N$ given above.

\begin{figure*}
    \includegraphics[width=1.0\textwidth]{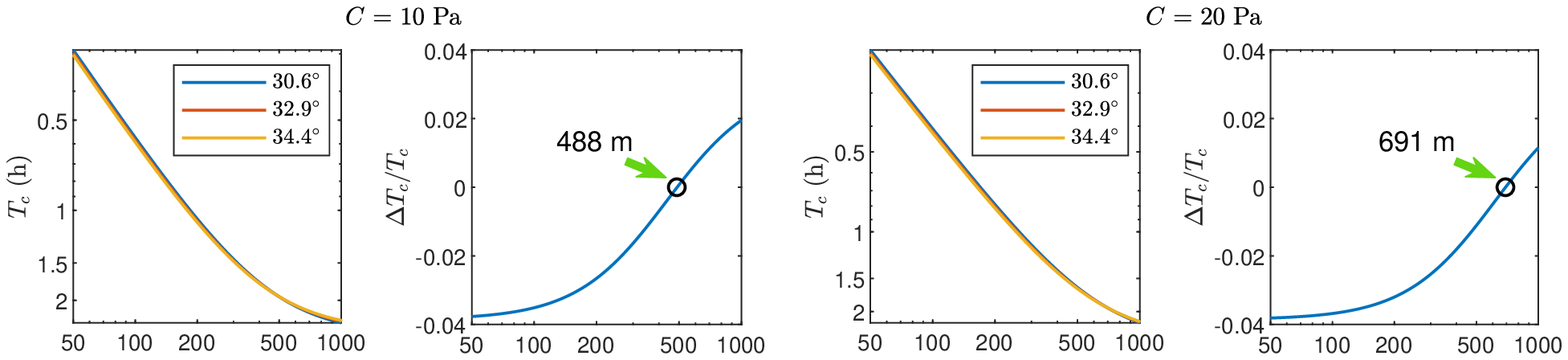}
    \caption{$T_c-D$ curves obtained with the continuum theory for $C$ = 10 Pa and 20 Pa and $\phi$ = 30.6$^\circ$, 32.9$^\circ$ and 34.4$^\circ$ ($\rho$ = 2.4 g/cm$^3$). The relative differences ($\Delta {T_c}/{T_c}$) shown in the 1st and 3rd panels are given as $[{T_c}(\phi  = 30.6^\circ ) - {T_c}(\phi  = 34.4^\circ )]/{T_c}(\phi  = 34.4^\circ )$. The marked black circles are points when $\Delta {T_c}$ = 0, and their horizontal coordinates (corresponding to the critical diameters when $\phi$ has no effect on $T_c$) are labeled in the figures.}
    \label{fig:fig_mus_Hol_diffC}
\end{figure*}

The $T_c-D$ curves for $C$ = 10 Pa and 20 Pa by applying the continuum theory are shown in Fig. \ref{fig:fig_mus_Hol_diffC} for $\phi$ = 30.6$^\circ$, 32.9$^\circ$, 34.4$^\circ$. From the figures, it follows that the friction angle has a minimum effect on $T_c$ at some critical diameter (denoted by $D_{cri,\phi}^A$). The values of $D_{cri,\phi}^A$ are illustrated in the figures, from which we find that $D_{cri,\phi}^A$ increases as $C$ increases. This trend is consistent with our numerical findings that $D_{cri,\phi}^N$ increases as $c$ increases. We also find that the values of $D_{cri,\phi}^A$ for $C$ = 10 Pa and 20 Pa are close to $D_{cri,\rho}^A$ given in Fig. \ref{fig:fig_density_Hol_diffC3}, with a difference of about 11\% for both cases.

\begin{figure}
    \includegraphics[width=0.47\textwidth]{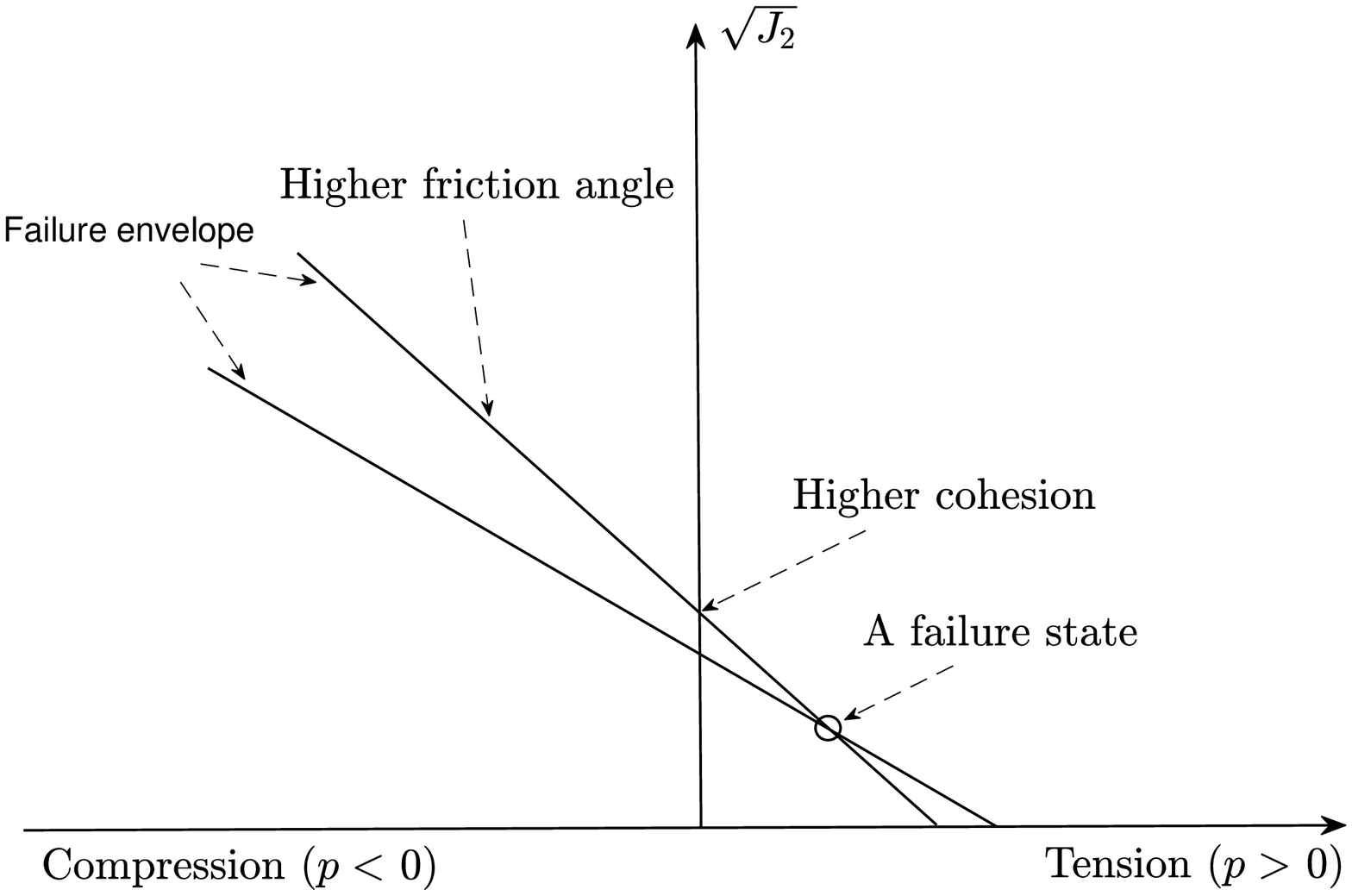}
    \caption{Illustration of the Drucker–Prager failure criterion applied to cohesive rubble piles \citep{holsapple2007spin}. Two failure envelopes with different friction angles are plotted. For a critically spinning rubble pile in tension state, a higher friction angle requires a higher cohesion to maintain the same failure state.}
    \label{fig:DP-envelope}
\end{figure}

By taking into account the close relationship between $\mu_S$ and $\phi$, we can expect that the effect of $\mu_S$ on the critical spin period obtained by our modeling is equivalent to the effect of $\phi$ predicted by the analytical model\footnote{In current work, we do not analyze the effect of friction angle resulting from the particle shape parameter $\beta$, which will be kept constant at 0.5 from Section 4.5 to the end.}. However, from Fig. \ref{fig:fig_mus_Hol_diffC}, we find that $T_c$ obtained by the analytical method increases as $\phi$ increases when $D$ < $D_{cri,\phi}^A$, which is opposite from the trend shown in our numerical outcomes when $D$ < $D_{cri,\phi}^N$. In fact, the critical diameter $D^A_{c,\phi}$ or $D^N_{c,\phi}$ corresponds to a state where the mean normal stress of a rubble pile is close to zero (see Fig. \ref{fig:fig_p_vs_D} and discussions in section 4.7). As shown in Fig. \ref{fig:DP-envelope}, the Drucker-Prager failure envelope has a larger slope for a higher friction angle. For a failure state located in the tension region, a higher friction angle indicates a higher cohesion.  Therefore, for a constant cohesion, the failure can be initiated with a larger friction angle.  However, in our SSDEM simulations, the structural stability is held by the interparticle contact network.  A large interparticle friction would guarantee a tougher structure and result in a smaller critical spin period $T_c$. This implies that the Drucker-Prager yield criterion may not be suitable to deal with this situation. Specially, caution is needed when applying the continuum theory to analyze the effect of friction angle.

\subsection{The ratio of \texorpdfstring{$c$}{Lg} to \texorpdfstring{$C$}{Lg}}
Due to the simplicity of the continuum theory, it has been widely used in the literature to predict the lower bound of internal strength of known SFRs under the assumption of rubble-pile structure \citep{rozitis2014cohesive,polishook20162,polishook2017fast}. Using our $T_c-D$ curves, it is interesting for us to compare our numerical results with the analytical results.

\begin{figure*}
    \includegraphics[width=1.0\textwidth]{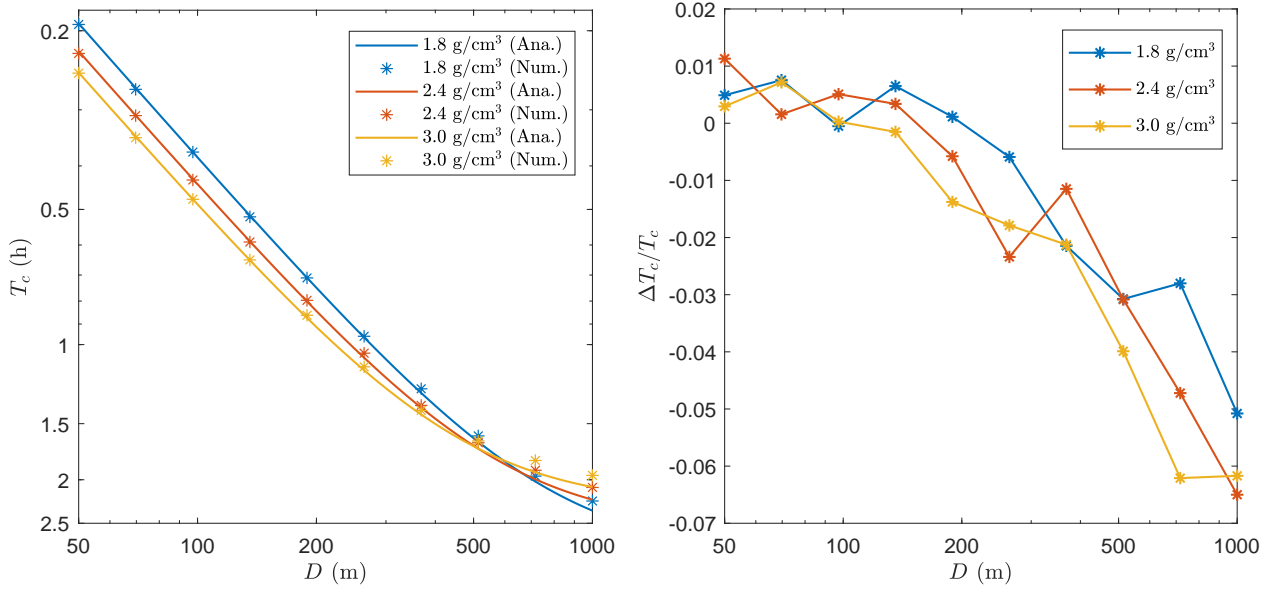}
    \caption{Left: the numerically obtained $T_c$ with respect to $D$ and the corresponding best-fit results for different densities using the analytical method. Right: the distributions of fitting errors ($\Delta {T_c}/{T_c}={[{T_c}({\rm{Num}}.) - {T_c}({\rm{Ana}}.)]/{T_c}({\rm{Num}}.)}$) with $D$. The parameters $\beta$ = 0.5, $\mu_S$ = 0.5, $c$ = 1600 Pa and the best-fit $C$ = 18.2 Pa are used.}
    \label{fig:fig_plot_fit}
\end{figure*}

In our model, the interparticle cohesion $c$ describes the microscopic strength due to the discrete nature of the SSDEM model while the bulk cohesion $C$ in the continuum theory reflects the macroscopic strength of the bulk body. The ratio of $c$ to $C$ is important to connect the two kinds of results together. We can use the analytical solution to fit our numerical results by tuning the value of $C$ to minimize the mean residual defined as
$$
{\rm{Residual}} = \frac{1}{{10}}\sum\limits_{i = 1}^{10} {\frac{{|{T_{{c_i}}}({\rm{Num}}.) - {T_{{c_i}}}({\rm{Ana}}.)|}}{{{T_{{c_i}}}({\rm{Num}}.)}}}
$$
The resulting best-fit $C$ (as well as the corresponding $k$, $c/k$ and $c/C$) for different sets of parameters are given in Table \ref{tab:fit}, from which we can see that the fits are quite robust, with the mean residual less than 3\%. The fitting results for different densities are shown in Fig. \ref{fig:fig_plot_fit} and we can see from the right panel that the maximum error is about 7\%. Specifically, the critical spin rates predicted by the analytical theory are generally smaller than the numerical results when the diameter is larger.

\begin{table*}
    \centering
    \caption{The best-fit value of $C$, $k$, $c/C$ and $c/k$ for each set of parameters (where friction angles are adopted from the spinup tests).}
    \label{tab:fit}
    \begin{tabular}{ccccccccccc}
    \hline
    NO. & $c$ (Pa) & $\beta$ & $\rho$ (g/cm$^3$) & $\mu_S$ & $\phi$ ($^\circ$) & $C$ (Pa) & $k$ (Pa) & $c/C$ & $c/k$ & Residual \\ \hline
    (1) & 800 & 0.5 & 2.4 & 0.5 & 32.9 & 9.1 & 10.8 & 87.9 & 74.3 & 2.7 \% \\
    (2) & 1600 & 0.5 & 2.4 & 0.5 & 32.9 & 18.1 & 21.4 & 88.4 & 74.7 & 2.0 \% \\
    (3) & 3200 & 0.5 & 2.4 & 0.5 & 32.9 & 35.9 & 42.5 & 89.1 & 75.3 & 1.5 \% \\
    (4) & 1600 & 0.5 & 1.8 & 0.5 & 32.9 & 18.1 & 21.4 & 88.4 & 74.7 & 1.6 \% \\
    (5) & 1600 & 0.5 & 3.0 & 0.5 & 32.9 & 18.2 & 21.5 & 87.9 & 74.3 & 2.2 \% \\
    (6) & 1600 & 0.5 & 2.4 & 0.3 & 30.6 & 14.6 & 17.5 & 109.6 & 91.6 & 2.8 \% \\
    (7) & 1600 & 0.5 & 2.4 & 0.7 & 34.4 & 19.5 & 22.9 & 82.1 & 69.5 & 1.6 \% \\ \hline
    \end{tabular}
\end{table*}

From the results of cases (1)-(5) in Table \ref{tab:fit}, it follows that both $c$ and $\rho$ have little influence on $c/k$ and $c/C$. Specifically, $c$ is proportional to the best-fit $C$ (note that the difference in $c/C$ among cases (1)-(3) is within 1\%). However, we find that the value of $\mu_S$ (or $\phi$) has a significant influence on $c/k$ and $c/C$, which should not be unexpected in view of the previous analysis already shows the inconsistent variation trend of $T_c$ with $\phi$ between the numerical and analytical results when $D<D_{cri,\phi}$.

For the cases with the nominal $\mu_S$ and $\beta$, our results give $c/C$ $\approx$ 88.3 and $c/k$ $\approx$ 74.7, which are taken as the mean value of results of cases (1)-(5) in Table. \ref{tab:fit}. In the work of \cite{zhang2018rotational}, they estimated $c/k$ $\approx$ 100, which is about 25\% higher than our results. This may arise from that we use different ways to define the critical spin period. Note that here we calculate $T_c$ according to the criterion that a rubble pile is globally destroyed (or globally deformed for larger bodies) while they measured $T_c$ based on the local failure region near the surface, which can make our allowable critical spin rates relatively higher and eventually reduce the ratio of $c/k$.

Based on the best-fit $c/k$ $\approx$ 74.7, Eq. \ref{eq:dpyc} is modified as
\begin{equation}
    \label{eq:dpyc_modify}
    \left\{ \begin{array}{l}
        \sqrt {{J_2}}  \le k' - 3s'p\\
        k' = \frac{c}{74.7},\;\;\;s' = 0.255
        \end{array} \right.
\end{equation}
which may be used to quickly judge whether a spinning rubble pile (with $\mu_S$ = 0.5 and $\beta$ = 0.5) attains global failure before performing a time-consuming simulation. From the right panel of Fig. \ref{fig:fig_plot_fit}, we can see that the fitting error is generally smaller when $D$ < $D_{cri,\rho}$ than $D$ > $D_{cri,\rho}$. Therefore Eq. \ref{eq:dpyc_modify} should be more applicable to $D$ < $D_{cri,\rho}$.

As shown in Fig. \ref{fig:Tc-diff_structure} and \ref{fig:Tc_diffN_diffD}, the unknown internal structure of a rubble pile can result in an uncertain $T_c$, with a level of a few percents, which results in a comparative level of uncertainty in the best-fit value of $c/C$ and $c/k$. We should accept this uncertainty and know that it is caused by the discrete nature of our SSDEM model. Fortunately, this uncertainty is insignificant and does not affect our main conclusions.

\subsection{The critical diameters}
The previous sections defined two critical diameters, $D_{cri,\rho}^{N}$ and $D_{cri,\phi}^{N}$, from the numerical results and two equivalent critical diameters, $D_{cri,\rho}^{A}$ and $D_{cri,\phi}^{A}$, with the analytical method, according to the characteristics of variation trends of $T_c$ with respect to $\rho$ and $\phi$, respectively. Using the obtained best-fit $C$, we are able to compare the critical diameters further.

For $c$ = 1600 Pa and 800 Pa, $D_{cri,\rho}^{N}$ and $D_{cri,\phi}^{N}$ have been calculated with our SSDEM simulations, as shown in Fig. \ref{fig:fig-density-num-oblate-800Pa} and \ref{fig:fig-mus-num-oblate-800Pa}. With the corresponding best-fit $C$ of 18.2 Pa and 9.1 Pa (given in Table \ref{tab:fit}), $D_{cri,\rho}^{A}$ and $D_{cri,\phi}^{A}$ can be obtained through the continuum theory by solving the equations
\begin{equation}
    \label{eq:partial_rho}
    \frac{{\partial {T_c}}}{{\partial \rho }} = 0
\end{equation}
and
\begin{equation}
    \label{eq:partial_phi}
    \frac{{\partial {T_c}}}{{\partial \phi }} = 0
\end{equation}
respectively, in which the expression for $T_c$ can be derived from Eq. \ref{eq:dpyc} (replace the inequality with an equal sign). Rather than deriving the complicated explicit analytical solutions of $D_{cri,\rho}^{A}$ and $D_{cri,\phi}^{A}$, we used numerical approach (use the central difference scheme like Eq. \ref{eq:chafenfangcheng}) to find the results. All the results are collected and shown in Table \ref{tab:Dc} for comparison.

\begin{table*}
    \centering
    \caption{The critical diameters obtained from our numerical simulations and the analytical method for the nominal oblate rubble piles ($\rho$ = 2.4 g/cm$^3$, $\beta$ = 0.5, $\mu_S$ = 0.5 and $\phi$ = 32.9$^\circ$).}
    \label{tab:Dc}
    \begin{tabular}{cccccc}
    \hline
    NO. & $c$ (Pa) & $C$ (Pa) & $D_{cri,\rho}$ (m) & $D_{cri,\phi}$ (m) & Method \\ \hline
    (1) & 1600 & 18.2 & 560 & 586 & Num. \\
    (2) & 1600 & 18.2 & 572 & 662 & Ana. \\
    (3) & 800 & 9.1 & 392 & 415 & Num. \\
    (4) & 800 & 9.1 & 406 & 469 & Ana. \\ \hline
    \end{tabular}
\end{table*}

We can see from Table \ref{tab:Dc} that, for $c$ = 1600 Pa and 800 Pa, the two $D_{cri,\rho}^N$ are very close to the corresponding $D_{cri,\rho}^A$, differing by 2.1\% and 3.5\%, respectively, and the differences between $D_{cri,\rho}^N$ and $D_{cri,\phi}^N$ are 4.5\% and 5.7\%, respectively. However, relatively larger differences are observed between $D_{cri,\phi}^N$ and $D_{cri,\phi}^A$, with differences of 12.1\% and 12.2\%, respectively.

Simply put, combining our numerical results and the analytical results, we find that the critical diameters $D_{cri,\rho}^N$, $D_{cri,\phi}^N$ and $D_{cri,\rho}^A$ are very close to each other, while a relatively larger difference is observed between $D_{cri,\phi}^N$ and $D_{cri,\phi}^A$. Given the opposite variation trend of $T_c$ with $\mu_S$ when $D$ < $D_{cri,\phi}^N$ (or $T_c$ with $\phi$ when $D$ < $D_{cri,\phi}^A$) between the two kinds of results, this discrepancy should not be unexpected. In view of the fact that the diameters, shapes, densities, and bulk cohesions of asteroids are usually unknown or have relatively large uncertainties (let alone the heterogeneous internal structure and cohesion distributions), the difference of $\sim$12\% is actually insignificant from a practical point of view.

\subsection{Compressive regime and tensile regime}
\begin{figure}
    \includegraphics[width=0.47\textwidth]{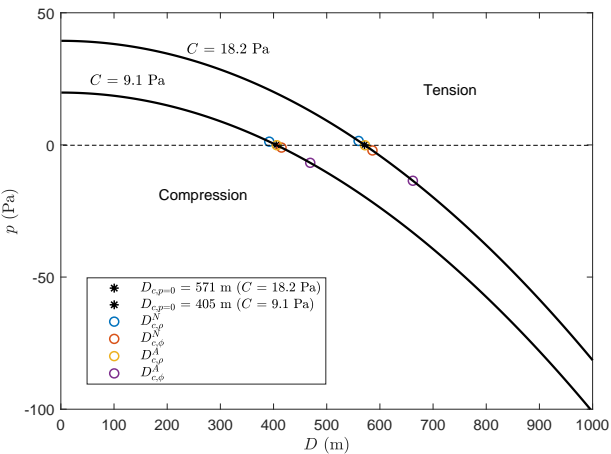}
    \caption{The change of mean normal stress $p$ with respect to $D$ for critically spinning rubble piles with $C$ = 18.2 Pa and 9.1 Pa (the nominal oblate shape, $\rho$ = 2.4 g/cm$^3$ and $\phi$ = 32.9$^\circ$ are used). The critical diameters calculated with the numerical and analytical methods are marked with circles.}
    \label{fig:fig_p_vs_D}
\end{figure}

Using Eq. \ref{eq:mean_normal_stress}, the mean normal stress $p$ of a spinning rubble pile can be easily calculated as a function of $D$ for a given bulk cohesion, shape, density and friction angle; this has been plotted in Fig. \ref{fig:fig_p_vs_D} for $C$ = 9.1 Pa and 18.2 Pa, for which the corresponding critical diameters are also labeled. The region below the $p$ = 0 line is in the compression state while the other is in tension. It is evident that $D_{cri,\rho}^N$, $D_{cri,\phi}^N$ and $D_{cri,\rho}^A$ are very close to the diameter when $p$ = 0 (denoted by $D_{cri,p=0}$), that is
\begin{equation}
    \label{eq:Dc_equal}
    {D_{cri,\rho}^N} \approx {D_{cri,{\phi}}^N} \approx {D_{cri,{\rho}}^A} \approx {D_{cri,p = 0}}
\end{equation}

\cite{holsapple2007spin} defined a ``gravity regime'' for larger bodies ($D$ > 10 km) and ``strength regime'' for smaller bodies ($D$ < 3 km) according to whether the gravity or tensile strength dominates in Eq. \ref{eq:dpyc}. Here we state that the critical diameters are relevant to the concepts of ``compressive regime'' ($p$ < 0) and ``tensile regime'' ($p$ > 0)\footnote{Note that the ``compressive regime'' and ``tensile regime'' defined here do not necessarily require any part inside the body to be in compression or tension.}. According to our numerical results, for a critically spinning rubble pile, $T_c$ decreases as $\rho$ increases in the compressive regime (as we generally expect) while the trend reverses when transitioning to the tensile regime. Moreover, we find that $\mu_S$ (or $\phi$) has a minimum effect on $T_c$ when the body is located at the separation between the two regimes. This can be understood as follows: according to the Drucker-Prager yield criterion of Eq. \ref{eq:dpyc}, the term $3sp$ $\equiv$ 0 for any $\phi$ once $p$ = 0, and even the whole equation can keep constant if we ignore the contribution of the $k$ term (note that $k$ also depends on $\phi$, but $k$ is usually insensitive to $\phi$).

From the results of Fig. \ref{fig:Tc-diff_structure} and \ref{fig:Tc_diffN_diffD}, we can see that the uncertainty of $T_c$ is only about 1\% at around $D_{cri,p=0}$, which indicates that the unknown internal structure has very little influence on the value of $D_{cri,\rho}^N$ and $D_{cri,\phi}^N$. Therefore the relationship of Eq. \ref{eq:Dc_equal} should still hold even considering the different particle arrangements or model resolutions.

At $p$ = 0, the critical diameter $D_{cri,p=0}$ can be found by solving the equations
\begin{equation}
    \left\{ \begin{array}{l}
        {{\bar \sigma }_x} + {{\bar \sigma }_y} + {{\bar \sigma }_z} = 0\\
        \frac{1}{6}\left[ {{{\left( {{{\bar \sigma }_x} - {{\bar \sigma }_y}} \right)}^2} + {{\left( {{{\bar \sigma }_y} - {{\bar \sigma }_z}} \right)}^2} + {{\left( {{{\bar \sigma }_z} - {{\bar \sigma }_x}} \right)}^2}} \right] = {k^2}
        \end{array} \right.
\end{equation}
By eliminating ${\bar \sigma }_y$, we have
\begin{equation}
    \label{eq:derive_p_0}
    \begin{array}{l}
        \bar \sigma _x^2 + {{\bar \sigma }_x}{{\bar \sigma }_z} + \bar \sigma _z^2 = {k^2}
    \end{array}
\end{equation}
By inserting Eq. \ref{eq:stress} into Eq. \ref{eq:derive_p_0} we can obtain the explicit expression of $D_{cri,p=0}$ in terms of $\rho$, $C$, $\phi$ and the shape (characterized by the ratios ${\gamma _i} = \frac{{2{a_i}}}{D}$, $i$ = 1, 2, 3), which can be complicated for a general body. However, for an oblate shape ($A_x$ = $A_y$ and $\gamma_1$ = $\gamma_2$), we have a much simpler expression
\begin{equation}
    \label{eq:Dc_phi}
    \begin{array}{l}
        {D_{cri,p=0}} = \frac{{\sqrt k }}{\rho }\sqrt {\frac{20}{{\sqrt 3 \pi G{A_z}\gamma _3^2}}}
    \end{array}
\end{equation}
that can be simplified further for our nominal oblate body with $\phi$ = 32.9$^\circ$,
\begin{equation}
    \label{eq:Dc_p_0_oblate}
    {D_{cri,p = 0}} = \left( {720\;{\rm{m}}} \right)\frac{{\sqrt {C/\left( {20\;{\rm{Pa}}} \right)} }}{{\rho /\left( {2\;{\rm{g/c}}{{\rm{m}}^3}} \right)}}
\end{equation}
For a general purpose to do a rough estimation, a spherical body with $\phi$ = 35$^\circ$ can be assumed and Eq. \ref{eq:Dc_p_0_oblate} needs to be modified slightly as
\begin{equation}
    {D_{cri,p = 0}} = \left( {695\;{\rm{m}}} \right)\frac{{\sqrt {C/\left( {20\;{\rm{Pa}}} \right)} }}{{\rho /\left( {2\;{\rm{g/c}}{{\rm{m}}^3}} \right)}}
\end{equation}
The above expressions demonstrate that the critical diameter scales with the square root of cohesion and inversely with the density. With the best-fit $c/C$, we are able to use these expressions to quickly calculate the critical diameters of cohesive rubble piles without running the simulations.

\begin{figure}
    \includegraphics[width=0.47\textwidth]{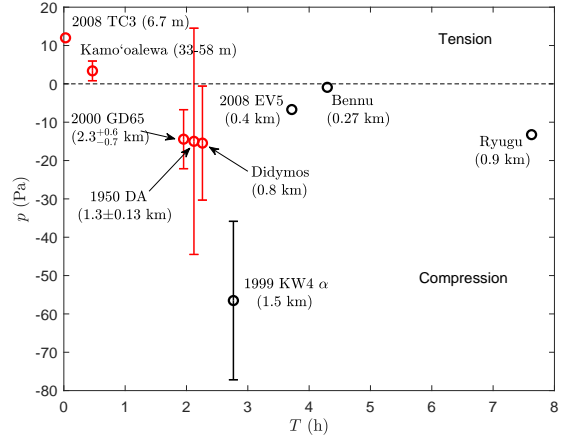}
    \caption{Illustration of mean normal stresses of 9 asteroids (note that the definitions of compressive and tensile regime do not necessarily require the body to be critically rotating). The asteroids marked with red circles are SFRs and the others are top-shaped asteroids with higher rotation periods. The errorbars are given due to their uncertain densities or diameters.}
    \label{fig:fig_p_of_real_asteroids}
\end{figure}

With known sizes, densities and spin rates, the mean normal stresses of 9 real asteroids in the solar system can be calculated with Eq. \ref{eq:mean_normal_stress}. The results are plotted as a function of their rotation periods and shown in Fig. \ref{fig:fig_p_of_real_asteroids}. The asteroids 2008 TC3, (469219) Kamo`oalewa (provisionally named as 2016 HO3), (60716) 2000 GD65, (29075) 1950 DA and (65803) Didymos \footnote{Actually, Didymos rotates with 2.26 h, which is slightly higher than the 2.2 h spin barrier presented in the begining of this paper. Note that this spin barrier is not definite and depends on density. Therefore, here we also simply take Didymos as a SFR for comparison.} are SFRs and the other four are top-shaped asteroids with high rotation periods. The results show that all of these asteroids are in the compressive regime, except for 2008 TC3 and Kamo`oalewa (1950 DA is more likely to be in the compressive regime based on the errorbar).

Of the two SFRs in tension, Kamo`oalewa is a ``quasi-satellite'' of Earth, with absolute magnitude $H$ of 24.3 and a rotation period of 28 min \citep{de2016asteroid}, which is also one of the two targets (the other one is the main-belt comet 133P) of a proposed Chinese mission. If assuming it is an S-type asteroid and the albedo $p_v$ is 0.1-0.3, we have $D$ = 33-58 m according to the relationship $D = \left( {1329\;{\rm{m}}} \right){10^{ - H/5}}/\sqrt {{p_v}}$. The LCDB database shows that the maximum amplitude of the lightcurve of Kamo`oalewa is 0.8. This gives $a_3/a_1$ = 1/2.1 if it is assumed to be a prolate body with the spin axis perpendicular to the observer. If Kamo`oalewa is a rubble pile, the minimum bulk cohesion is 2-11 Pa according to Eq. \ref{eq:dpyc} (assuming $\rho$ = 1-2.4 g/cm$^3$ and $\phi$ = 35$^\circ$). Obviously, the level of this cohesive strength is not significantly higher than what we generally think asteroids have. Accordingly, it is impossible to constrain its internal structure only by the cohesion. If the mission succeeds in the future, Kamo`oalewa will be the first object to be directly checked whether an asteroid in the tensile regime can be a rubble pile.

\section{Discussion and Conclusions}
In this work, with a cohesion-enhanced SSDEM method, we performed a series of numerical simulations to model the spin-up process of small cohesive self-gravitating rubble piles ranging from 50 m to 1,000 m in diameter. The critical spin periods $T_c$ of the rubble piles were calculated and the dependencies of $T_c$ on the interparticle cohesion $c$, the shape parameter $\beta$, the bulk density $\rho$ and the static friction coefficient $\mu_S$ were investigated. Specifically, we explored how these dependencies can change with the bulk diameter $D$, and compared our results to the analytical solution derived from the continuum theory due to \cite{holsapple2007spin}.

Assuming a nominal oblate shape with $a_3/a_1$ = 0.9, the critical spin periods were determined over a varied parameter space with the SSDEM simulations by finding the moment when the moment of inertia changes by 1\% during the spinup. Unlike for large rubble piles, our results indicate that both the interparticle cohesion and the shape parameter in our SSDEM model can greatly influence the critical spin rates of small rubble piles. At diameters when the gravity is ignorable compared to the centrifugal force at the critically spinning state, we found that $T_c$ is proportional to $\frac{{D\sqrt \rho  }}{{\beta \sqrt c }}$, which is consistent with the predictions given by the analytical method (the interparticle cohesion $c$ is proportional to the bulk cohesion $C$, as shown in Section 4.5), except that our model also considers the contribution of the contact area $\beta$.

The curves of $T_c$ as a function of $D$ were obtained with the numerical simulations and the results were fitted with the analytical method. The ratio of $c$ to $C$ was obtained following the fitting, and we found that this ratio remains constant for different cohesions and bulk densities, while it strongly depends on the friction angle (or $\mu_S$). For the nominal parameters $\mu_S$ = 0.5 and $\beta$ = 0.5 ($\phi$ = 32.9$^\circ$), the average value of $c/C$ is about 88.3, with an uncertainty of a few percents for varied particle arrangements and model resolutions.

The numerical results of $T_c$ with respect to $\rho$ demonstrate that $T_c$ increases with $\rho$ at larger bulk diameter and then this trend reverses at a critical diameter $D_{cri,\rho}$ as $D$ decreases. It was shown that this phenomenon can be also found with the continuum theory. This fact implies that a rubble pile with a higher density does not necessarily achieve a greater bulk strength, especially for smaller bodies. For situations when the gravity is ignorable, the minimum bulk cohesion of a spinning rubble pile can be approximately calculated according to Eq. \ref{eq:Tc_simple_nominal}
\begin{equation}
    \label{eq:C_minimum_for_small_D}
    C = \frac{{2\rho {D^2}}}{{T_c^2}}
\end{equation}
for which $C$ is proportional to $\rho$. In the solar system, typical bulk densities of C, S and M type asteroids are 1.4, 2.69 and 4.7 g/cm$^3$, respectively \citep{britt2002asteroid}. Accordingly, this result implies that small M type fast spinning rubble-pile asteroids need 2.4 times larger minimum cohesion to keep the body intact than a C type under the same conditions, and it predicts that small M type rubble-pile SFRs are more difficult to survive than C type ones. However, we still do not know whether any correlation exists between the regolith cohesion and material composition of asteroids, and the current poor knowledge of the physical characteristics of small asteroids does not allow to test this prediction.

The effect of static friction coefficient $\mu_S$ on $T_c$ was also explored with the SSDEM method. We found that greater $\mu_S$ (changed from 0.3 to 0.7) always strengthens the bulk bodies. However, our numerical outcomes revealed that $\mu_S$ has a minimum effect on $T_c$ at a critical diameter $D_{cri,\phi}$ close to $D_{cri,\rho}$. With the continuum theory, we found similarly that the friction angle has a minimum effect on $T_c$ at a critical diameter. However, the dependency of $T_c$ with $\phi$ in the two methods is opposite from each other when $D$ < $D_{cri,\phi}$. This is a remarkable difference between our numerical results and the analytical results, which reminds us that caution is needed when using the continuum theory for small cohesive rubble piles, especially when considering the effect of friction angle. Frankly speaking, currently it is not allowable for us to judge which results are more reliable or more close to real cohesive rubble-pile asteroids (of course our numerical results are more consistent with expectations). Here we encourage other researchers to do similar simulations to test the $T_c-\phi$ dependency with a different cohesion-included SSDEM code (or other modeling methods), and perform comparisons with our results and the analytical results.

Fortunately, since we usually only care about ``gravel"-like material in asteroid research field, which corresponds to a narrow friction angle range of about 30$^\circ$-40$^\circ$\citep{lambe1969soil}. We can calculate with Eq. \ref{eq:dpyc} that the fluctuation of $C$ due to the range is typically less than 25\% at $D$ < $D_{cri,\phi}$. Since the difference between our numerical results and the analytical results at $D$ < $D_{cri,\phi}$ is very small (see the right panel of Fig. \ref{fig:fig_plot_fit}), we can still safely use the continuum theory to roughly predict the minimum bulk cohesion a small rubble-pile asteroid needs to hold its structure.

Another interesting aspect that we note from the results is that the numerically obtained $D_{cri,\rho}^N$ and $D_{cri,\phi}^N$, and the analytically obtained $D_{cri,\rho}^A$, are close to the diameter $D_{cri,p=0}$ at which the mean normal stress equals zero, except that a slightly greater difference (about 12\%) is observed between $D_{cri,\phi}^N$ and $D_{cri,\phi}^A$. Note that $D_{cri,p=0}$ is the separation between the compressive regime and tensile regime; this fact may imply that different mechanical characteristics exist in the two regimes, respectively. Also, we can derive a simple analytical expression for $D_{cri,p=0}$, as shown in Eq. \ref{eq:Dc_p_0_oblate}, which can be used to calculate the value of $D_{cri,\rho}^N$ and $D_{cri,\phi}^N$; these critical diameters are useful for understanding the dynamical behavior of a spinning cohesive rubble pile.

\begin{figure}
    \includegraphics[width=0.47\textwidth]{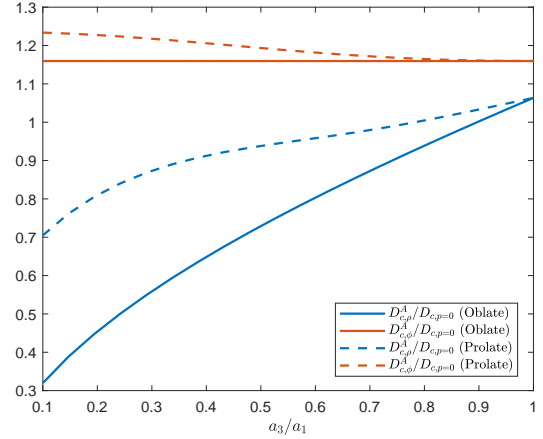}
    \caption{The ratios of $D_{cri,\rho}^A$ to $D_{cri,p=0}$ and $D_{cri,\phi}^A$ to $D_{cri,p=0}$ calculated with the analytical method for different shape (oblate and prolate with different $a_3/a_1$). $\phi$ = 32.9$^\circ$ is assumed.}
    \label{fig:fig_Dc_ratio_32_9}
\end{figure}

However, according to Eq. \ref{eq:dpyc}, we find that the relationship of diameters $D_{cri,\rho}^A$, $D_{cri,\phi}^A$ and $D_{cri,p=0}$ (see Eq. \ref{eq:Dc_equal}) is not always close, but strongly depends on the shape (as well as the friction angle). The ratios of $D_{cri,\rho}^A/D_{cri,p=0}$ and $D_{cri,\phi}^A/D_{cri,p=0}$ are shown in Fig. \ref{fig:fig_Dc_ratio_32_9} for oblate and prolate rubble piles with different axis ratio $a_3/a_1$, from which we can see that $D_{cri,\rho}^A$ increases as $a_3/a_1$ increases and tends to equal $D_{cri,p=0}$ at $a_3/a_1$ $\sim$ 0.9, which happens to be the nominal value used in this work. So, though this relationship is possibly not a general conclusion (which needs more simulations to verify the relationships for different shapes and friction angles), in view of the fact that the adopted nominal parameters and shape are good representatives of real rubble-pile asteroids, the relationship is still meaningful and the simple expression of $D_{cri,p=0}$ is useful to give a good estimation of $D_{cri,\rho}^N$ and $D_{cri,\phi}^N$ for a given cohesive rubble pile.

\section*{Acknowledgements}
The simulations in this work were carried out at the University of Maryland on the yorp cluster administered by the Department of Astronomy and the deepthought2 supercomputing cluster administered by the Division of Information Technology. This work is financially supported by the National Natural Science Foundation of China (Grant NOs. 11873098, 11661161013, 11673072, 11633009, 11761131008), CAS Interdisciplinary Innovation Team, the Strategic Priority Research Program on Space Science, the Chinese Academy of Sciences, Grant No. XDA15020302) and Foundation of Minor Planets of the Purple Mountain Observatory. Y. Z. acknowledges funding from the Universit\'e C\^ote d'Azur ``Individual grants for young researchers'' program of IDEX JEDI. We also appreciate Professor Keith Holsapple for his insightful discussions.

\section*{Data availability}
The data underlying this article is modeled and generated with the software $pkdgrav$, which can be shared on reasonable request to the coauthor Prof. Derek C. Richardson (dcr@astro.umd.edu). The simulated data can be shared on reasonable request to the corresponding author.



\bibliographystyle{mnras}

\bibliography{mybibfile}



\bsp	
\label{lastpage}
\end{document}